\documentclass[twocolumn,prd,aps,superscriptaddress,preprintnumbers,tightenlines,showpacs,nofootinbib,eqsecnum,amsfonts,amsmath]{revtex4}
\pdfoutput=1

\usepackage{color}
\usepackage{calc}
\usepackage{amsmath,amssymb,graphicx}
\usepackage{tensor}
\usepackage{bm}
\usepackage{times}
\usepackage[varg]{txfonts}
\usepackage{float}
\usepackage{dcolumn}

\usepackage[abs]{overpic}
\usepackage{pict2e}

\usepackage[caption=false]{subfig} 

\DeclareMathAlphabet{\mathcalstd}{OMS}{cmsy}{m}{n}
\DeclareMathAlphabet{\mathpzc}{OT1}{pzc}{m}{it}

\newcommand{\chiIMR}{\chi_\text{IMR}}
\newcommand{\chiPN}{\chi_\text{PN}}

\newcommand{\LNh}{\hat{\mathbf{L}}_N}

\newcommand{\blambda}{\bm{\lambda}}
\newcommand{\bLambda}{\bm{\Lambda}}
\newcommand{\bchia}{\bm{\chi}_a}
\newcommand{\bchis}{\bm{\chi}_s}
\newcommand{\chis}{\chi_s}
\newcommand{\chia}{\chi_a}
\newcommand{\chiadL}{\bchia \cdot \LNh}
\newcommand{\chisdL}{\bchis \cdot \LNh}

\newcommand{\Mchirp}{\mathcal{M}_\text{c}}
\newcommand{\MM}{\mathcal{M}}

\newcommand{\LIGOCaltech}{LIGO Laboratory, California Institute of Technology, 
Pasadena, CA 91125, USA}
\newcommand{\TAPIR}{Theoretical Astrophysics, California Institute of Technology, 
Pasadena, CA 91125, USA}
\newcommand{\UIB}{Departament de F\'isica, Universitat de les Illes Balears, 
Crta. Valldemossa km 7.5, E-07122 Palma, Spain}
\newcommand{\Cardiff}{School of Physics and Astronomy, Cardiff University, Queens Building, CF24 3AA, Cardiff, United Kingdom}
\newcommand{\ICTS}{International Centre for Theoretical Sciences, Tata Institute of Fundamental Research, IISc Campus, Bangalore 560012, India}

\begin{document}


\title{Testing the validity of the single-spin approximation in inspiral-merger-ringdown waveforms}

\author{Michael P\"urrer}
\affiliation{\Cardiff}

\author{Mark Hannam}
\affiliation{\Cardiff}

\author{P. Ajith}
\affiliation{\ICTS}
\affiliation{\LIGOCaltech}
\affiliation{\TAPIR}

\author{Sascha Husa}
\affiliation{\UIB}

\begin{abstract}
Gravitational-wave signals from black-hole binaries with non-precessing spins are 
described by four parameters -- each black hole's mass and spin. It has been shown that 
the dominant spin effects can be modeled by a \emph{single} spin parameter, 
leading to the development of several \emph{three-parameter} waveform models. 
Previous studies indicate that 
these models should be adequate for gravitational-wave detection. In this paper we focus 
on the systematic biases that would result from using them to estimate 
binary parameters, and consider a one-parameter family of configurations at one choice of mass
ratio and effective single spin. We find that for low-mass binaries within that family of configurations, 
where the observable waveform is dominated by the inspiral, the systematic bias in all physical parameters is 
smaller than 
the parameter uncertainty due to degeneracies between the  
mass ratio and the spins, at least up to signal-to-noise ratios (SNRs) of 50. For
higher-mass binaries, where the merger and ringdown make a greater contribution to
the observed signal, the bias in the mass ratio is comparable to its uncertainty at
SNRs of only $\sim$30, and the bias in the measurement of the total spin is 
\emph{larger} than the uncertainty defined by the 90\% confidence region even at an SNR of 
only 10. Although this bias may be mitigated in future models by a better choice of 
single-effective-spin parameter, these results suggest that it may be possible to
accurately measure \emph{both} black-hole spins in intermediate-mass binaries.  
\end{abstract}


\maketitle


\section{Introduction} 
\label{sec:introduction}

The inspiral and merger of black-hole and neutron-star binaries are the most promising
sources for the first direct detection of gravitational waves (GWs) with the Advanced LIGO 
(aLIGO) and Virgo (AdV) detectors~\cite{Harry:2010zz,aVIRGO,Abadie:2010cf}, 
and are expected to provide a wealth of 
astrophysical information (see, e.g.,~\cite{Sathyaprakash:2009xs}). The optimal technique to locate
their signals in the detector data is to cross-correlate the data against a large bank of theoretical 
signal templates. A search across the full parameter space of component masses,
spins, sky locations, orientations and the distance is computationally extremely challenging. 
Partly for this reason searches in data from the initial LIGO and Virgo 
detectors~\cite{Aasi:2012rja,Colaboration:2011np,Abadie:2011kd,Abadie:2010yb} 
focussed on binaries with non-spinning components, for which a two-dimensional template 
bank suffices, greatly reducing the computational cost. The two dimensions are defined by 
combinations of the component masses, and the effects of the sky-location, orientation
and distance on the signal observed by a single detector can be absorbed into an overall 
amplitude scale factor. 

Such simplifications are not possible for generic spinning binaries, where 
the components' spins cause precession of the orbital plane and of the spins themselves, leading to far more complex
GW signals. However, if we consider only spins aligned/anti-aligned to the binary's orbital angular 
momentum, then the only spin effects are on the inspiral rate and the signal amplitude -- 
the basic waveform structure is unchanged from the non-spinning case. Including the aligned/anti-aligned
spin effects in the waveform templates makes it possible to detect a much larger volume of the 
binary parameter space, including in some cases a significant fraction of precessing 
binaries~\cite{Ajith:2009bn,Ajith:2011ec,Brown:2012qf}. It is also possible that non-precessing-binary
models can be used as the basis for constructing generic waveform models~\cite{Schmidt:2012rh,Pekowsky:2013ska}.
Note that this study analyzes only the $l=2, m=\pm 2$ modes of the gravitational wave signal.

The inclusion of the (non-precessing) black-hole spins doubles the dimensionality of the 
search parameter space over non-spinning searches. However, studies of inspiral 
dynamics using post-Newtonian expansions, and of merger and ringdown with numerical
solutions of Einstein's equations, show that the dominant spin effects can be 
modeled with a single parameter~\cite{Ajith:2009bn,Ajith:2011ec,Santamaria:2010yb}. 
 This has motivated the development of waveform
models parametrized by only the binary's mass ratio and effective total spin (the binary's 
total mass appears as a simple overall scale 
factor)~\cite{Vaishnav:2007nm,Reisswig:2009vc,Ajith:2009bn,Ajith:2011ec,Santamaria:2010yb}.
Also, recent work on implementing non-precessing-spin template banks has exploited the 
partial degeneracy between the two spins~\cite{Brown:2012qf,Ajith:2012mn}. 

The use of a single effective spin parameter is also motivated by the high computational 
cost of fully general relativistic numerical simulations. Work to date on phenomenological
waveform models suggests that we require at least four simulations in each direction of 
parameter space that we wish to model. A model of the full seven-dimensional parameter
space of generic binary waveforms would require $4^7 \approx 16,000$ simulations, which
are not feasible before the commissioning of aLIGO and AdV~\cite{Ajith:2007qp,Ajith:2007kx,Ajith:2007xh,Ajith:2009bn, Santamaria:2010yb,Hannam:2009vt}. The most ambitious study to
date includes ``only'' $\sim$200 waveforms, at moderate mass ratios and black-hole 
spins~\cite{Mroue:2013xna}. It is therefore important that we exploit any degeneracies
that reduce the dimensionality of the parameter space that we must model. 

While single-effective-spin models are believed to capture the phenomenology
of non-precessing-binary signals with sufficient fidelity for GW detection, little is known about 
how well they would perform if used to estimate the source parameters following a 
detection. 
The single-spin approximation is only valid in the leading-order post-Newtonian spin terms
(although it holds to higher order when both masses are equal, and for extreme mass
ratios, where the influence of the smaller black hole's spin is negligible),
and does not hold through merger, where the appropriate single spin becomes the
total spin angular momentum of the two black 
holes~\cite{Buonanno:2007sv,Rezzolla:2007rd,Tichy:2008du,Lousto:2009mf}. 
If we use a single-effective-spin waveform model
for parameter estimation, what will be the bias in the measurement of the black-hole masses,
and of the spin parameter itself? Obviously, if we approximate the two black-hole spins with 
a single spin, then we cannot use this model to measure the individual black-hole spins; on 
the other hand, if a single spin parameter models the dominant spin effects, then both spins will be 
difficult to measure even if we \emph{did} use a double-spin model. We will return to this
point later. 

In this paper we explore the parameter biases due to the use of a single-effective-spin model. Since
we expect the single-effective-spin approximation to become less valid for higher mass ratios, we consider
a set of configurations at the highest mass ratio of the numerical simulations that were used to calibrate
current phenomenological models, 1:4~\cite{Ajith:2009bn,Santamaria:2010yb}. 
The computational cost of numerical simulations
precludes an exhaustive study, so we focus on one value of the effective spin parameter, 
$\chi_{\rm IMR} := (m_1\chi_1 + m_2\chi_2) / (m_1 + m_2) = 0.45$. (Here $m_1$ and $m_2$ are the masses, 
and $\chi_1$ and $\chi_2$ are the Kerr parameters of the black holes). We produce a set of five simulations with differing values of 
the individual black-hole spins ($\chi_1$ and $\chi_2$), but with the same value of $\chi_{\rm IMR}$. From the 
numerical-relativity waveforms we construct hybrid PN-NR waveforms, which are in turn 
compared against one of the phenomenological models, ``IMRPhenomC'' 
(see Sec.~\ref{sec:Phenom-models} for a 
more detailed description of the waveform model). By identifying the IMRPhenomC waveform that
agrees best with each hybrid, we estimate the parameter biases due to the use of a 
single-effective-spin model. 

There are a number of issues that make it difficult to draw conclusions from this procedure. 
The results will be skewed by artifacts in the construction of the particular waveform model that we
use (the details of the phenomenological ansatz, the coverage of the parameter space by numerical
waveforms, and the accuracy of the waveforms), which may swamp the errors due to the 
single-effective-spin approximation. Previous studies have shown that the main source of uncertainty 
in hybrid PN-NR waveforms
is in the PN regime~\cite{Hannam:2010ky,MacDonald:2011ne,Boyle:2011dy,Ohme:2011zm}, 
and as such our results will depend on the PN approximant we use in 
our hybrids, and on the hybridization frequency. We discuss these issues further, and the
steps we have taken to mitigate them, in Sec.~\ref{sec:IMR-results}.

The layout of the paper is as follows. We summarize the single-spin approximation, waveforms models 
and our numerical waveforms in Secs.~\ref{sec:Phenom} and \ref{sec:NRwaveforms}. In Sec.~\ref{sec:PN-results}
we make a preliminary study of biases in the inspiral regime, where we can compare
single- and double-spin PN models using the same PN approximant, and do not have to concern
ourselves with issues of hybridization or phenomenological modeling. In addition to quantifying the
parameter biases due to the single-effective-spin approximation for low-mass binaries (for mass-ratio 1:4
and moderate spins), this section also provides context and contrast to the full inspiral-merger-ringdown
results, which are in Sec.~\ref{sec:IMR-results}. 

 
\section{Preliminaries}
\label{sec:Phenom}

\subsection{The single-spin approximation}
\label{sec:single-spin-approximation}

We consider black-hole binaries where the spins are aligned or anti-aligned with the orbital angular 
momentum. Then the spins and the angular momentum do not precess, which leads to a 
considerable simplification of the GW signal over generic configurations. These aligned-spin 
waveforms are parametrized by the black-hole masses and spins. 

A single effective spin parameter $\chiIMR := (m_1\chi_1 + m_2\chi_2) / M$ has been used in the 
construction of the non-precessing phenomenological inspiral-merger-ringdown (IMR) models 
presented in Refs.~\cite{Ajith:2009bn,Santamaria:2010yb}. These models parametrize the 
waveforms by their mass $M$, symmetric mass ratio $\eta = m_1 m_2 / M^2$, and the 
effective spin parameter $\chiIMR$. They incorporate a PN description of the inspiral, 
while the merger and 
ringdown regimes are tuned using the results of numerical simulations.
A recent study~\cite{Ajith:2011ec} has addressed how well a related ``reduced spin'' 
parameter motivated by PN theory works for inspiral searches. This PN model has 
been shown to be sufficiently accurate for GW searches (``effectual''), and to agree
well with the full two-spin waveforms (``faithful'') when either the spins or the 
masses are equal.

In constructing the PN reduced-spin parameter, we note that all spin effects can be 
described by two parameters $\chis \equiv \chisdL \equiv (\chi_1+\chi_2)/2$ and 
$\chia \equiv \chiadL \equiv (\chi_1-\chi_2)/2$, which remain constant throughout the evolution. 
The dimensionless spin parameters $\chi_i$ are defined as $\chi_i = S_i / m_i^2$, where 
$S_i$ is the spin of black hole $i$. The leading order spin term due to spin-orbit coupling 
appearing at 1.5PN order in the amplitude and the phase can be represented by a single 
``reduced spin'' parameter (see e.g.~\cite{Ajith:2011ec,Poisson:1995ef})
\begin{equation}
	\label{eq:chi_PNdef}
	\chiPN \equiv \chis + \delta \chia -\frac{76\eta}{113} \chis,
\end{equation}
where $\eta = m_1 m_2 / M^2$ and $M=m_1 + m_2$.

In contrast, the ``effective spin'' parameter used in the phenomenological models for black-hole 
binaries with non-precessing spins~\cite{Ajith:2009bn,Santamaria:2010yb} is defined as a 
simple mass-weighted linear combination of the spins
\begin{equation}
	\label{eq:chi_IMRdef}
	\chiIMR \equiv \left(m_1 \chi_1 + m_2 \chi_2 \right) / M = \chis + \delta \chia.
\end{equation}
For equal masses both spin parameters are a function of the symmetric combination of the 
spins $\chis$ only. The only difference is an overall factor. 
For unequal masses both spin parameters depend on the symmetric 
and anti-symmetric spin combinations. The difference between the spin parameters depends 
linearly on $\eta$ and therefore goes to zero for infinite mass-ratio.

Historically, the non-precessing phenomenological IMR models to date have used the effective 
spin parameter $\chiIMR$ defined in Eq.~\eqref{eq:chi_IMRdef} due to its simple form.
While this choice was sufficient to build effectual non-precessing waveform models, we will 
present evidence that suggests that $\chiPN$ is a better choice and should be used for future models.

\subsection{Phenomenological single-spin models}
\label{sec:Phenom-models}

We can quantify the agreement between families of waveforms with the same value of $\chiPN$ 
or $\chiIMR$, but in order to estimate the parameter bias that would result from the single-spin 
approximation, individual waveforms are not sufficient; we require a waveform family. In this study
we compare our PN/NR hybrids with the phenomenological model for black-hole binaries with 
non-precessing spins presented in~Ref.~\cite{Santamaria:2010yb}. For consistency with 
the labeling used within the LIGO-Virgo Collaboration~\cite{LAL} we refer to this model as ``IMRPhenomC''. (``IMRPhenomA'' refers to a model of nonspinning 
binaries~\cite{Ajith:2007qp,Ajith:2007kx,Ajith:2007xh}, and ``IMRPhenomB'' to an earlier model of 
non-precessing binaries~\cite{Ajith:2009bn}; we choose to use ``IMRPhenomC'' because it 
incorporates higher-order PN information in the inspiral phasing, but also make cross-checks 
against the IMRPhenomB model.)

The model waveforms are parametrized by their total mass $M = m_1 + m_2$, symmetric mass 
ratio $\eta = m_1 m_2 / M^2$, and the effective total spin parameter $\chiIMR$ defined in 
Eq.~\eqref{eq:chi_IMRdef}.
The waveform is represented in the Fourier domain as $h(f) = A(f) e^{i \Psi(f)}$. 
The amplitude $A(f)$ and phase $\Psi(f)$  are modeled separately. 
The IMRPhenomC amplitude is constructed from two parts: a PN inspiral amplitude with the 
addition of a higher order frequency term, and a ringdown portion, both of which are fit to the 
model hybrids. For the inspiral portion of the phase IMRPhenomC uses the complete 
TaylorF2~\cite{Sathyaprakash:1991mt,Cutler:1994ys,Droz:1999qx,Ajith:2012az} PN inspiral phasing 
(up to 3.5PN order, although the spin terms are complete 
only up to 2.5PN). Only the late inspiral/merger phase is fitted in a narrow frequency range 
$[0.1 f_\text{RD}, f_\text{RD}]$ to numerical simulations, while the ringdown waveform is 
obtained from analytically derived quasi-normal mode expressions for the frequency and 
attached continuously to the merger phase. For both the amplitude and the phase smooth 
$\tanh$-window functions are used to connect the individual parts.

The model is a power series in the frequency $f$, and the coefficients in the model are written 
as polynomials in the two physical parameters $\eta$ and $\chiIMR$ (the total mass is an overall 
scale factor), and it is the coefficients of these polynomials that are then calibrated to hybrids of 
PN and NR waveforms. There are 45 free parameters in IMRPhenomC, although the final model 
is a function of only $\{M,\eta,\chiIMR\}$. The hybrids used to construct IMRPhenomC were produced 
in the frequency domain, using TaylorF2 for the PN part and a rather broad fitting window 
$M f \in [0.01, 0.02]$. The construction of frequency-domain PN/NR hybrids is discussed in more 
detail in Sec.~\ref{sec:hybridization}.

\subsection{Matches, fitting factors and confidence regions} 
\label{sec:matches}

We quantify the agreement between two waveforms, $h_1(f)$ and $h_2(f)$, with the standard 
inner product weighted by the power spectral density $S_n(f)$ of a detector~\cite{Cutler94}, 
called the \emph{overlap}
\begin{equation} 
  \langle h_1 | h_2 \rangle = 4 \, \text{Re} 
  \int_{f_\text{min}}^{f_\text{max}}
  \frac{\tilde h_1(f) \tilde h_2^\ast(f)}{S_n(f)} df.  
\end{equation}
The inner product is calculated in terms of the frequency-domain waveforms $\tilde{h}(f)$. 
The frequency range in which the detector is deemed sensitive is $[f_{\rm min}, f_{\rm max}]$.
Let $\hat h(f) \equiv \tilde h(f) / \sqrt{\langle h | h \rangle}$ be the normalized frequency-domain 
waveform. The match between two normalized waveforms is then defined as their inner product, 
maximized over time and phase shifts of the waveform, 
\begin{equation} 
  \text{M}(h_1,h_2) =  \max_{\Delta t, \Delta \phi} \langle \hat h_1 \, | \, \hat h_2 \rangle.
\end{equation}

Given a signal waveform $h(\blambda)$ with physical parameters $\blambda$ and a template 
$x(\bLambda)$ with physical parameters $\bLambda$ we define the \emph{fitting factor}
\begin{equation}
  \text{FF} = \max_{\Delta t, \Delta \phi, \bLambda} \langle \hat x(\bLambda) \, \vert \, \hat h(\blambda) \rangle.  
\end{equation}
Instead of the fitting factor we will often quote the fully optimized mismatch 
\begin{equation}
  \label{eq:mismatch_opt}
  \MM = 1-\text{FF}.
\end{equation}

The match quantifies the physical agreement between two waveforms (since the time of arrival and overall
phase of the waveform do not change the underlying physics of the binary). The fitting factor is a measure of
how well a matched-filter search with a given waveform family can perform in detecting a particular signal;
a fitting factor greater than 0.965 indicates that no more than 10\% of signals will be lost in a search. It does
not tell us, however, how well the parameters of the best-match template will agree with the true source 
parameters of the signal.

A PN-NR hybrid binary waveform that has been produced for a given total mass $M$ can be trivially rescaled 
to a different mass. Therefore, the match between two such waveforms can be optimized over the total mass. 
The phenomenological model ``IMRPhenomC'' used in this study depends in addition on the symmetric 
mass-ratio $\eta$ and the effective spin $\chiIMR$, which allows us to compute fitting factors by optimizing 
matches over $\bLambda = \{M,\eta,\chi\}$.

In this paper we compare PN-NR hybrids signals with ``IMRPhenomC'' with reference to the expected sensitivity 
of the Advanced LIGO detector~\cite{Abbott:2007kv,Shoemaker:aLIGO,2010CQGra..27h4006H}.  
Early science runs are expected around 2015~\cite{Aasi:2013wya}. At its optimum sensitivity several years later, 
the anticipated sensitivity is given by the ``zero-detuned high-power'' noise curve~\cite{T0900288}. 
We use a linear interpolation of this expected PSD and choose $f_{\rm min} = 15$\,Hz, and $f_{\rm max} = 8$\,kHz.
 
For PN matches we choose the upper frequency of the overlap integral as the frequency of the 
innermost stable 
circular orbit (ISCO) of a test particle around a Schwarzschild black hole 
$f_\text{ISCO} = v^3_\text{ISCO}/(\pi M)$, 
where $v_\text{ISCO} = 1/\sqrt{6}$ just as in~\cite{Ajith:2011ec}. The Schwarzschild ISCO is an 
arbitrary point at which to terminate the PN waveform, but it corresponds to the choice commonly 
made in detector searches~\cite{Babak:2012zx}. 

The model parameters for the waveform that best matches the signal correspond to the parameters
that are most likely to be recovered in a GW measurement. We are also interested in the range of
parameters that would be recovered in 90\% of observations at a given SNR, i.e., the 90\% 
confidence region for that SNR, which illustrates the statistical uncertainty in the measurement. 

At high SNRs the confidence region can be estimated by Fisher-matrix
methods~\cite{Finn:1992xs,Cutler:1994ys,Poisson:1995ef,Arun:2004hn}, 
while in general one should construct the full posterior probability distribution 
function~\cite{Sluys:2008a, Sluys:2008b, Veitch:2010, Feroz:2009,Aasi:2013jjl}. 
The latter is computationally very expensive, but Ref~\cite{Baird:2012cu} shows that 
it is possible to produce a good approximation to the correct confidence region by computing matches
between the model waveform with the physical parameters of the signal, and model waveforms with a 
range of neighboring parameters. All neighbouring waveforms that have a match greater than some
threshold are within the 90\% confidence region. The threshold for a given SNR $\rho$ assuming a 
3-dimensional parameter space is~\cite{Baird:2012cu} 
\begin{equation}
  \label{eq:CR}
  M ( h_\text{m}(\theta), h_\text{m}(\theta_0) ) \ge 1 - 3.12 / \rho^2, 
\end{equation}
where $\theta$ are arbitrary waveform
parameters (in this case $M$, $\eta$, and $\chiPN$ or $\chiIMR$), and $\theta_0$ are the correct
parameters, and $h_\text{m}(\theta)$ are the model waveforms. 

We have computed fitting factors and the associated best parameters with two different methods. 
The Nelder-Mead Amoeba~\cite{Nelder:1965zz} simplex method has been used to compute fitting factors for a 
range of masses (see Sec.~\ref{sec:IMR-FF-IMRPhenomC}). For selected masses we have computed $90\%$ 
confidence regions by sampling the matches on a suitably fine grid in $(\Mchirp, \eta, \chi)$ space. The latter 
computation is a lot more expensive, but more reliable --- in some cases the amoeba calculation can be trapped 
in a local minimum, especially when the confidence region in question is not simply connected. At low masses 
and high SNRs confidence regions can be very elongated filaments and a transformation to rotate and squash 
the region into a more compact form is then helpful to keep the computation within a reasonable cost. This is 
related to the alternative parameter-space coordinates that are being used in placing waveforms in search template
banks~\cite{Ohme:2013nsa,Tanaka:2000xy,Brown:2012qf,Pai:2012mv}.


\section{Numerical waveforms} 
\label{sec:NRwaveforms}

To fully test the single-spin approximation across the binary parameter space, we would need to 
perform, for each of a wide range of choices of $\eta$ and $\chiPN$ (or $\chiIMR)$, a series of 
simulations for choices of different black-hole spins that correspond to the same value of $\chiPN$ 
($\chiIMR$). In doing so, we would have produced enough simulations to construct a complete two-spin waveform model --- but the high computational cost of doing so is one of the motivations for 
producing a single-effective-spin model in the first place! 

We expect the single-effective-spin approximation to become less accurate as the binary mass ratio increases, 
and so in this study we focus on the largest mass ratio that was considered in the numerical simulations 
used to calibrate current phenomenological models~\cite{Ajith:2009bn,Santamaria:2010yb}, 
$q = m_2/m_1 = 4$. 
We choose an effective total spin of 
$\chiIMR = 0.45$; this is a relatively large total effective spin for which we can also choose a wide 
range of individual black-hole spins. 

In our simulations we set the total mass $M=1$ and using the convention $m_1 < m_2$ have $m_1 = 0.2$ 
and $m_2 = 0.8$. We then have $\chiIMR = 0.2 \chi_1 + 0.8 \chi_2$. With our choice of $\chiIMR = 0.45$ 
we let $\chi_1$ (the spin of the smaller BH), vary between -0.75 and +0.75. (Due to the large 
junk-radiation content 
in Bowen-York initial data for highly spinning binaries, we do not consider spins higher than 0.75.) 
Along this line of $\chiIMR=0.45$ in the $(\chi_1,\chi_2)$ plane we pick configurations for 
$\chi_1 = -0.75, -0.25, 0.25, 0.75$. In addition we also add the configuration with equal spins 
$\chi_1=\chi_2=0.45$. 
The latter configuration is important since IMRPhenomC~\cite{Santamaria:2010yb} assumes equal 
spins for its PN part. 
Our chosen configurations are summarized in table~\ref{tab:configs} (also see Fig.~\ref{fig:plots_chi_plane}).

\begin{table}[h] 
  \begin{tabular}{lcccccccc}
    \hline\hline
      Run & $\chi_1$ & $\chi_2$ & $m_1 / h_\text{min}$ & Cycles & $D/M$ & $e_{\phi,GW}$ & $M_f$ & $a_f/M$\\
      \hline
      1 & -0.75 & 0.75 	& 44.4 & 29 & 10.739 & 0.0003	& 0.966 &	0.84\\
      2 & -0.25 & 0.625 & 38.5 & 28 & 10.782 & 0.0006	& 0.969 &	0.79\\
      3 & 0.25 	& 0.5 	& 38.5 & 28 & 10.831 & 0.0007	& 0.971 &	0.74\\
      4 & 0.45 	& 0.45 	& 38.5 & 28 & 10.853 & 0.0027 & 0.973 &	0.72\\
      5 & 0.75 	& 0.375 & 44.4 & 28 & 10.889 & 0.0014	& 0.972 &	0.68\\
      \hline\hline
    \end{tabular}
    \caption{The series of $q=4$, $\chiIMR=0.45$ configurations used in this study 
    (also see Fig.~\ref{fig:plots_chi_plane}). We show the spin parameters of 
    the individual black holes, the resolution on the finest grid level with respect to the smallest black hole, 
    $m_1 / h_\text{min}$, the number of GW cycles before the amplitude maximum at merger, 
    the initial separation $D/M$, the eccentricity $e_{\phi,GW}$ measured from the GW phase~\cite{Purrer:2012wy} 
		and the final mass and spin. For the equal-spin configuration only one step of eccentricity reduction was performed.
}
    \label{tab:configs}
\end{table}

The numerical simulations were performed with the BAM 
code~\cite{Brugmann:2008zz,Husa:2007hp}, which evolves 
black-hole-binary puncture initial data~\cite{Brandt:1997tf,Bowen:1980yu} 
(generated using a pseudo-spectral elliptic solver~\cite{Ansorg:2004ds}), 
and evolves them with the $\chi$-variant of the
moving-puncture~\cite{Campanelli:2005dd,Baker:2005vv,Hannam:2006vv}
version of the BSSN~\cite{Shibata:1995we,Baumgarte:1998te} 
formulation of the 3+1 Einstein evolution equations. 
Spatial finite-difference derivatives are sixth-order accurate in the 
bulk~\cite{Husa:2007hp}, Kreiss-Oliger dissipation terms converge at fifth order, 
and a fourth-order Runge-Kutta algorithm is used for the time evolution. 
The gravitational waves emitted by the binary are calculated from the
Newman-Penrose scalar $\Psi_4$, and the details of our implementation of
this procedure are given in \cite{Brugmann:2008zz}. 
	
The basic grid setup for this study is based on a setup used for a convergence 
series of $q=4$, $\chi_1 = \chi_2=0.75$ simulations. In general, the mass-ratio 
has the most significant impact on choosing a grid setup, since the apparent 
horizons (AH) of the black holes must be resolved by the finest grid level. 
We have found that choosing the size of the innermost box about a factor 
$1.5 \times$ the size of the AH of the smaller BH leads to good 
accuracy. Since the spin of the smaller BH varies in the five configurations 
considered in this study, the individual grid setup is tuned differently for each 
simulation. We found that sufficient accuracy
could be achieved by using a minimum of 16 buffer zones (rather than the 
formal requirement of 32) between mesh-refinement boxes.
The Courant factor has been reduced to $C=0.4$ (from $C=0.5$ in our previous
work) to curb the contribution of 
time-integration error to the overall NR error. A detailed study of the accuracy of 
these simulations, and the effects of the errors due to both the spatial 
finite-differencing and the timestepping, will be presented in a forthcoming 
paper~\cite{Husa:2013}. 

We have performed two iterations of the eccentricity reduction method detailed in
Ref.~\cite{Purrer:2012wy} for each of the configurations. The final eccentricities are below 0.0015. 
The numerical waveforms are obtained by extracting the $l=m=2$ mode of $\psi_4$ at $r=180M$. 
	
While we have not performed convergence tests for the NR waveforms, we expect that
the results are robust because we have used a ``safe'' grid setup that has led to accurate 
results for related aligned-spin configurations with q=4 and $\chi_1 = \chi_2=0.75$. 
In Sec.~\ref{sec:IMR-FF-IMRPhenomC} we show how the numerical resolution 
affects fitting factors and biases of the $(0.45, 0.45)$ waveform with IMRPhenomC, and 
find that it does not affect our overall results.

\subsection{Construction and accuracy of PN-NR hybrid waveforms} 
\label{sec:hybridization}

The NR waveforms we have computed for this study do not cover the full aLIGO 
sensitivity band for lower-mass binaries.
Therefore we need to hybridize the NR waveform with a PN approximant.
Different choices are possible for constructing such PN-NR 
hybrids (see Ref.~\cite{Ajith:2012az} for a summary). 
We wish to compare our IMR hybrids with the IMRPhenomC model, in which the inspiral is modeled
 by the TaylorF2 
frequency-domain PN approximant. In order to minimize effects arising from differing PN approximants,  
we choose to create hybrids with TaylorF2.

We use a frequency domain hybridization method as described in~\cite{Santamaria:2010yb}. We include the NR $\psi_4$
waveform data from the time immediately after the passage of the burst of initial junk radiation, up to the point where
the ringdown is dominated by numerical noise, and apply a Planck tapering window~\cite{McKechan:2010kp} of width $300$ M
at the start of the dataset. The waveform is further padded with zeroes before computing the FFT to increase the
frequency resolution. The NR strain is calculated in the Fourier domain as $\tilde h_\text{NR}(f) =
\tilde\psi_4^\text{NR}(f) / (2\pi f)^2$. The matching procedure aligns $\tilde h_\text{PN}(f;t_0,\phi_0) = \tilde
h_\text{F2}(f) + 2\pi f t_0 + \phi_0$ and $\tilde h_\text{NR}(f)$ by a least squares fit over a fitting interval
$[f_1,f_2]$, which we discuss in more detail below. We then determine the matching frequency $f_m \in [f_1,f_2]$ at which
the NR and PN phases coincide by a root-finding algorithm. The PN and NR amplitudes are aligned separately without any
freedom to adjust parameters.

After having settled on an approximant and choice of hybridization method we can still choose the frequency region over
which the hybridization is performed. Intuitively it makes sense to hybridize at as low a frequency as possible so as to
extract as much useful information from the NR data as we can, and to minimize errors in the PN approximant, which
increase with GW frequency. 
It also stands to reason that the matching interval $[f_1,f_2]$ should not be too narrow as the fit would then be 
prone to pick up spurious oscillations in the NR data. For the hybrids considered in this study we have chosen an
interval length of $M\Delta \omega \sim 0.01$ and a matching frequency of about $M\omega_m \sim 0.07$. This results 
in a relative matching width of $\Delta \omega / \omega_m \sim 0.14$ and is consistent with the choice advocated in Ref.~\cite{MacDonald:2011ne}.

One accessible measure to gauge the quality of a hybrid are the parameter errors given by the least squares fit. At first
glance a high quality fit appears to be very desirable. However, the PN model used in the fit gives a worse approximation
of the NR phase as we go to higher frequencies and thus small standard errors in the fitting parameters do not
necessarily imply that the hybrid will be very faithful.

In a practical sense it is useful to think of the difference between hybrids with varying hybridization regions 
(assuming reasonable comparison interval, i.e. not unreasonably high frequencies) as a way of quantifying 
the error in the hybrid caused by the PN and NR data. We would like to know how such variations in the 
hybrid construction manifest themselves in biases and uncertainties. 
In Sec.~\ref{sec:IMR-FF-IMRPhenomC} we verify that our results are robust with respect to 
hybridization artifacts.


\section{Results for PN single-effective-spin models} 
\label{sec:PN-results}

A single-effective-spin PN model has recently~\cite{Ajith:2011ec} been shown to be an effective search
template (i.e., fitting factors $>0.97$) as well as ``faithful'' to two-spin signals (i.e., non-optimized matches are
also $>0.97$), when either the spins or the masses are equal. Here we address the question of 
biases and uncertainties incurred by the single-effective-spin approximation. While there is a 
preferred ``reduced-spin'' parameter $\chiPN$ in the PN regime, we also generalize the model used 
in Ref.~\cite{Ajith:2011ec} to arbitrary definitions of an effective spin parameter and compare with a 
model built from $\chiIMR$, the effective spin parameter used by current phenomenological 
waveform models.

The construction of a single-effective-spin model is straightforward. We choose one based on the 
TaylorF2 approximant. The model is based on the mapping 
$(\chi_1,\chi_2,\eta;f) \mapsto (\chi_\text{eff},\eta;f)$, where 
$\chi_\text{eff} = \chi_\text{eff}(\chi_1,\chi_2,\eta)$ is an arbitrary effective spin parameter. 
To build the model an inverse of this mapping is needed which requires a relation between 
$\chi_1$ and $\chi_2$. We choose to use only the symmetric part of the input spins 
(i.e., setting $\chia=0$) and define the frequency domain single spin model strain as
\begin{equation}
	\tilde h_\text{M}(\chi_\text{eff}(\chi_1,\chi_2,\eta),\eta;f) := \tilde h_\text{F2}(\chis,\chis,\eta;f).
\end{equation}
With this definition the model represents equal spin configurations exactly. For the choice $\chi_\text{eff} = \chiPN$ this model is identical to the one defined in Ref.~\cite{Ajith:2011ec}.

\begin{figure}[t]
	\centering
	\includegraphics[height=3in]{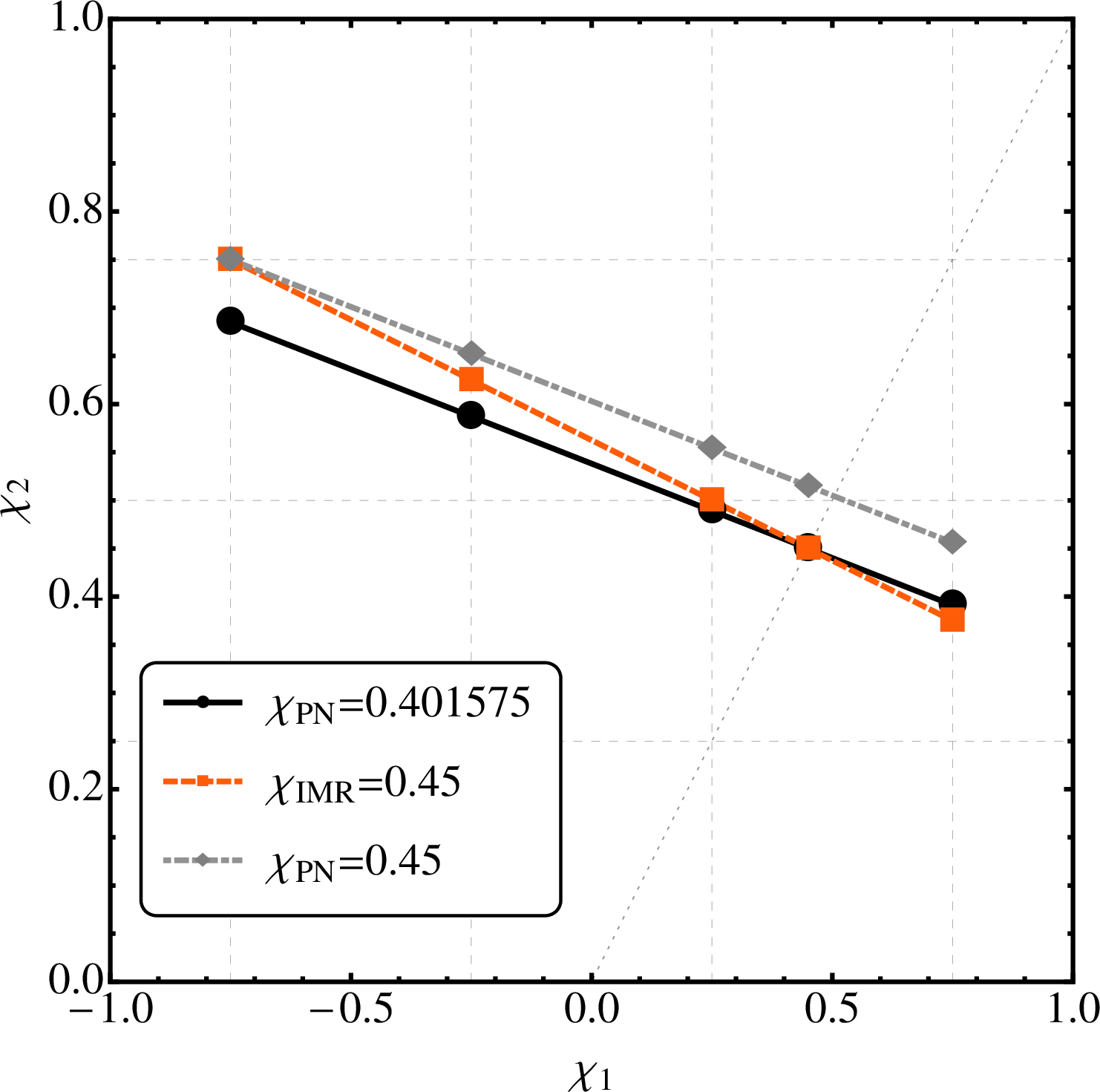}
	\caption{Configurations on lines of constant ``reduced'' spin parameter $\chiPN$ and 
	``effective'' spin parameter $\chiIMR$ chosen in this study for mass-ratio $q=4$. Two of 
	these configurations lie on the line $\chi_1=\chi_2$ (thin dotted line).}
  \label{fig:plots_chi_plane}
\end{figure}

We choose the following mass-ratio $q=4$ configurations for the comparison of single spin PN 
models (see Fig.~\ref{fig:plots_chi_plane}). For the $\chiIMR$ model we select the same cases 
along $\chiIMR=0.45$ (orange, dashed) as used for NR simulations. The configurations for the 
$\chiPN$ model are chosen along a line $\chiPN = 0.401575$ (black solid), rather than 
$\chiPN=0.45$ (gray, dot-dashed). The reason is that $\chiPN = 0.401575$ intersects 
$\chiIMR=0.45$ at the equal spin $(\chi_1,\chi_2)=(0.45, 0.45)$ configuration, for which both 
models are exact. For $\chi=0.45$ and $\chi_1=-0.75$ the symmetric combination of the spins 
$\chis$ vanishes and thus $\chiIMR=\chiPN$ at this point. The largest deviation between the spin 
parameters happens at the largest positive spin of the smaller BH $\chi_1=0.75$ and is about $20 \%$.

\begin{figure}[htbp]
  \centering
    \includegraphics[width=0.5\textwidth]{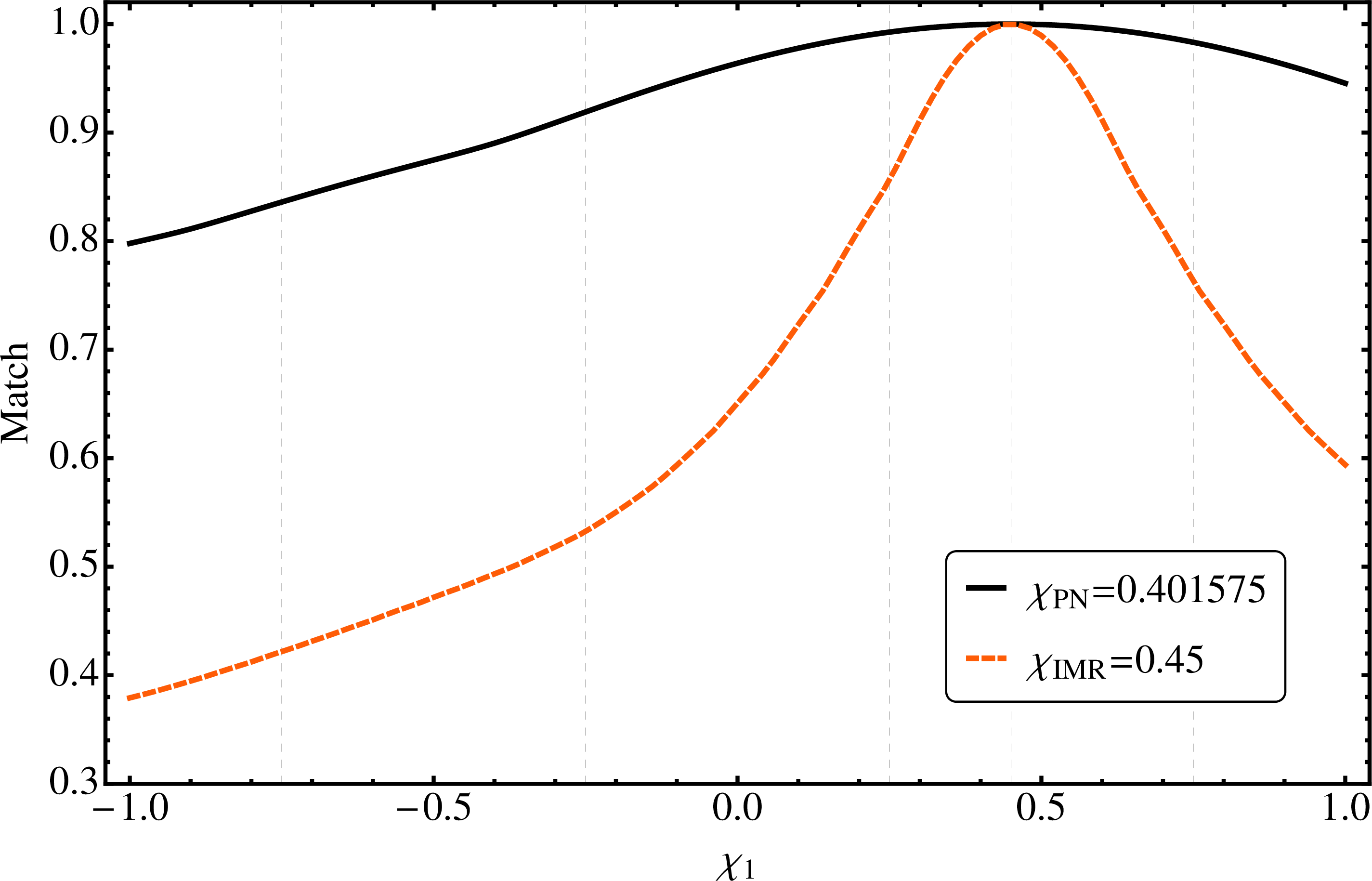}
  \caption{Matches of TaylorF2 signal waveforms along lines of $\chi_\text{eff}=const$ with single spin PN models with parameters $\chiIMR=0.45$ or $\chiPN=0.401575$ at $7 M_\odot$. The signal waveforms lie on lines $\chiIMR=0.45$ or $\chiPN=0.401575$, respectively.
}
  \label{fig:plots_PN-model-1p-matches}
\end{figure}

It is well known that $\chiPN$ is the (almost) optimal single-effective-spin parameter in the PN 
regime (this combination appears explicitly in the leading-order spin-orbit 
coupling)~\cite{Ajith:2011ec,Poisson:1995ef}.
The superiority of $\chiPN$ over $\chiIMR$ is illustrated in Fig.~\ref{fig:plots_PN-model-1p-matches} by how quickly matches between a single spin model (based on either $\chiPN$ or $\chiIMR$) and TaylorF2 signals (again at constant $\chiPN$ or $\chiIMR$, respectively) degrade when one moves away from the point $\chi_1=\chi_2$, where the models are exact, along a line of the respective $\chi_\text{eff}=const$.

As an example we compute fully optimized matches (fitting factors) and parameter biases for 
TayorF2 signal waveforms chosen as above with each of the single spin models for a system 
mass of $M=7M_\odot$. To quantify the statistical uncertainty we also calculate $90\%$ confidence 
regions in the 3-dimensional space of model parameters $(\Mchirp,\eta,\chi)$.

From an astrophysical point of view, a compact binary with $M=7 M_\odot$ could correspond to 
an actual NS-BH binary with component masses $(1.4,5.6)$. The NS would be expected to 
have very small spin which is contrary to some of the configurations chosen here. However, 
our goal is to compare how well the single spin approximation works in the PN and IMR regimes
(see Sec.~\ref{sec:IMR-FF-IMRPhenomC}) and therefore we choose the same range of spin
values for both PN and IMR models.

The results are summarized in Tabs.~\ref{tab:PNchiPN_FF_biases_7MS} and~\ref{tab:PNchiIMR_FF_biases_7MS}. 
The fully optimized mismatches (see Eq.~\eqref{eq:mismatch_opt}) have been computed from $90\%$ confidence 
regions at SNR 50 which will be discussed later. We have performed a least-squares fit to the elongated 
direction of the filament-like confidence regions and subsequently carried out a local minimization starting from the 
best match found along the curve fit. Although the knowledge of the confidence regions is not needed to compute 
fitting factors this method leads to more reliable results than simpler optimization methods.

We define biases as $\Delta \Lambda := \Lambda_\text{recovered} - \Lambda_\text{true}$,
where $\Lambda$ is one of the model parameters $(\Mchirp,\eta,\chi)$. Note that in tabs.~\ref{tab:PNchiPN_FF_biases_7MS}
and~\ref{tab:PNchiIMR_FF_biases_7MS} we have dropped the subscript PN/IMR in the relative bias of the spin parameters.
For ease of comparison we also give the absolute bias in the spin.

\begin{table}[h] 
  {\footnotesize
		\newcolumntype{.}{D{.}{.}{-1}}
    \begin{tabular}{lc....}
      \hline\hline
	    \multicolumn{1}{c}{Case $(\chi_1,\chi_2)$} & 
			\multicolumn{1}{c}{Mismatch $\MM$} & 
			\multicolumn{1}{c}{$\Delta \Mchirp/\Mchirp [\%]$} &
	    \multicolumn{1}{c}{$\Delta \eta /\eta [\%]$}  & 
			\multicolumn{1}{c}{$\Delta \chi /\chi [\%]$} &
			\multicolumn{1}{c}{$\Delta \chiPN$} \\
	    \hline
			(-0.75, 0.685104) 	& $6\times 10^{-4}$		& 0.07 	& -13.7 	& 13.4  & 0.054\\
			(-0.25, 0.587144) 	& $2\times 10^{-4}$		& 0.04 	& -8.8 		& 8.5 	& 0.034\\
			(0.25,  0.489184) 	& $2\times 10^{-5}$		& 0.01 	& -2.8 		& 2.7   & 0.011\\
			(0.45,  0.45)				& $0							$		& 0 		& 0		 		& 0     & 0\\
			(0.75,  0.391224) 	& $4\times 10^{-5}$		& -0.01 & 1.3 		& -2.0  & -0.008\\
     \hline\hline
    \end{tabular}
  }
  \caption{Fully optimized mismatches and biases between single spin PN model using $\chiPN$ and TaylorF2 signals 
  along $\chiPN=0.401575$ for $7M_\odot$.  
	}
    \label{tab:PNchiPN_FF_biases_7MS}
\end{table}

\begin{table}[H] 
  {\footnotesize
		\newcolumntype{.}{D{.}{.}{-1}}
	  \begin{tabular}{lc....}
	  	\hline\hline
	    \multicolumn{1}{c}{Case $(\chi_1,\chi_2)$} & 
			\multicolumn{1}{c}{Mismatch $\MM$} & 
			\multicolumn{1}{c}{$\Delta \Mchirp/\Mchirp [\%]$} &
	    \multicolumn{1}{c}{$\Delta \eta /\eta [\%]$}  & 
			\multicolumn{1}{c}{$\Delta \chi /\chi [\%]$} &
			\multicolumn{1}{c}{$\Delta \chiIMR$} \\
	    \hline
			(-0.75, 0.75) 	& $5\times 10^{-4}$	& 0.09 	& -20.5 	& 25.2 	& 0.113\\
			(-0.25, 0.625) 	& $2\times 10^{-4}$ & 0.06 	& -14.1 	& 16.4  & 0.074\\
			(0.25,  0.5) 		& $3\times 10^{-5}$ & 0 		& 1.1 		& 2.2  	& 0.01\\
			(0.45, 	0.45) 	& 0					& 0 		& 0 			& 0     & 0\\
			(0.75,  0.375) 	& $5\times 10^{-5}$	& 0 		& -0.5 		& -3.9 	& -0.018\\
	    \hline\hline
	   \end{tabular}
  }
  \caption{Fully optimized mismatches and biases between single spin PN model using $\chiIMR$ and TaylorF2 signals along
 	$\chiIMR=0.45$ for $7M_\odot$.
	}
    \label{tab:PNchiIMR_FF_biases_7MS}
\end{table}

The fitting factors for both single-effective-spin models are very high, with the fully optimized mismatch 
below $0.1 \%$. The models are exact at the equal-spin configuration and therefore the true 
parameters are recovered. As we move away from the equal-spin configuration the mismatch 
becomes larger and the parameter biases increase.
The bias in the chirp mass $\Mchirp = M \eta^{3/5}$ is overall very small, below $0.1\%$, consistent with
standard results. 
This is expected since the leading factor in the PN phase evolution for non-precessing binaries is 
proportional to $1/(\Mchirp \pi f)^{5/3}$, which  is dominated by the chirp 
mass~\cite{Finn:1992xs,Arun:2004hn,Baird:2012cu}. 
In contrast, there is considerable bias in the spin parameter and symmetric mass-ratio $\eta$ for the 
very unequal spin configurations. For the $\chiPN$-model the modulus 
of the biases in $\eta$ and $\chi$ increases to about $15\%$ for the configuration with $\chi_1=-0.75$, 
which is the farthest from the equal-spin case. The biases are worse for the $\chiIMR$-model and 
reach about $23\%$ for the $\chi_1=-0.75$ configuration. The absolute spin bias is at most 
$\Delta\chi \sim 0.05$ for $\chiPN$, while it rises to twice that value for $\chiIMR$. 

We know from PN theory that $\chiPN$ provides a better single-effective-spin approximation at low
masses, but these results quantify the difference in parameter biases in using either $\chiPN$ or
$\chiIMR$. The bias resulting from the use of $\chiIMR$ is up to twice as large as that due to using 
a model parametrized by $\chiPN$. 

At first glance, the results in Tabs.~\ref{tab:PNchiPN_FF_biases_7MS} and 
\ref{tab:PNchiIMR_FF_biases_7MS}
suggest that the single-effective-spin approximation is entirely inappropriate for parameter estimation:
the uncertainty in the mass ratio and the spin parameter can be as high as $\sim 15$\%, even if 
we use the $\chiPN$ approximation. This in turn suggests that, if a single-effective-spin approximation
behaves poorly, then we may be able to accurately resolve the \emph{individual} black-hole spins
with a complete two-spin model. Before making this conclusion, we should consider the 
statistical uncertainty in the parameter measurement for likely aLIGO and AdV SNRs. 

\begin{figure}[htbp]
	\centering
		\hspace*{-1.2mm}\includegraphics[width=0.5\textwidth]{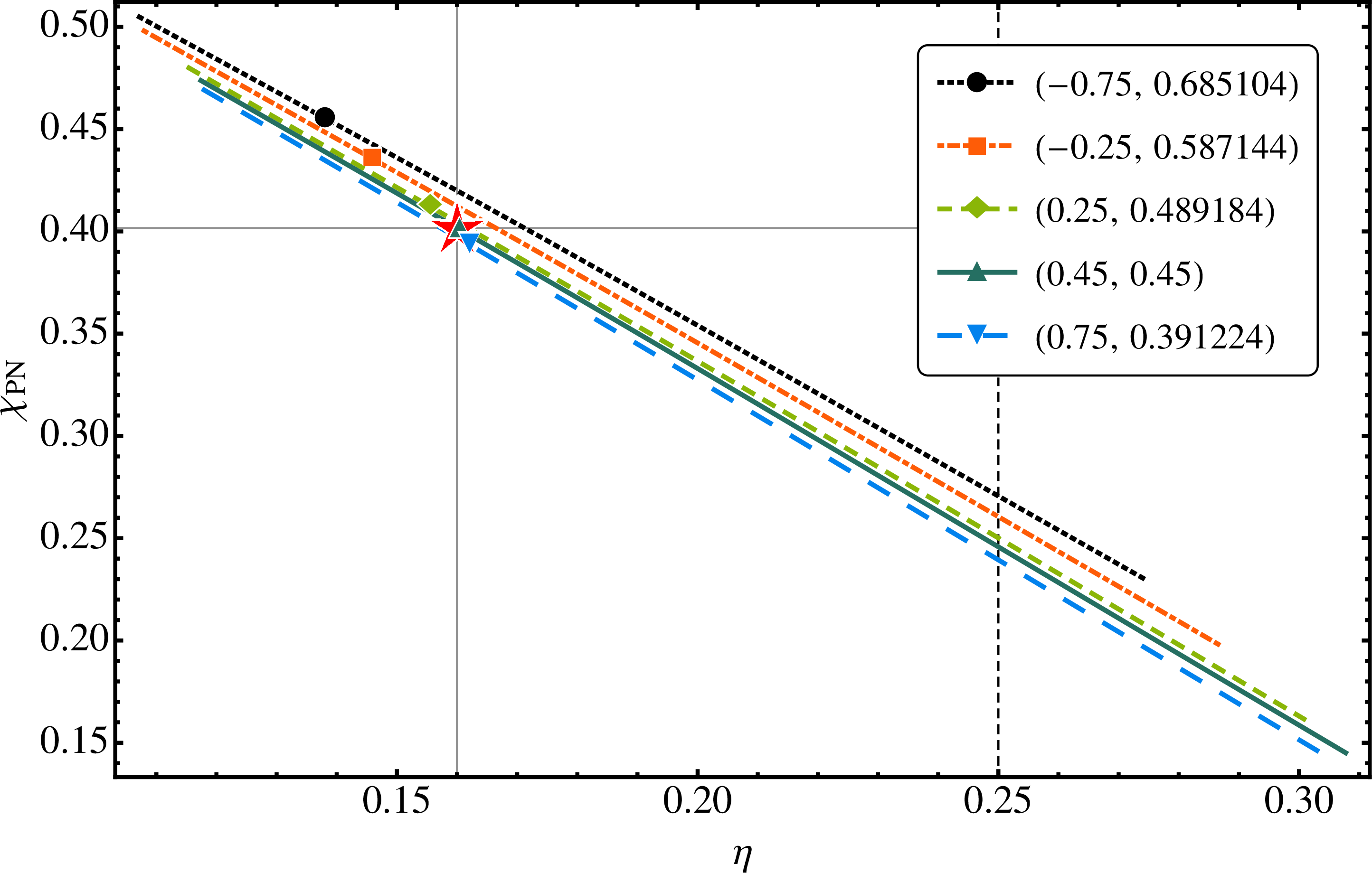}
		\includegraphics[width=0.493\textwidth]{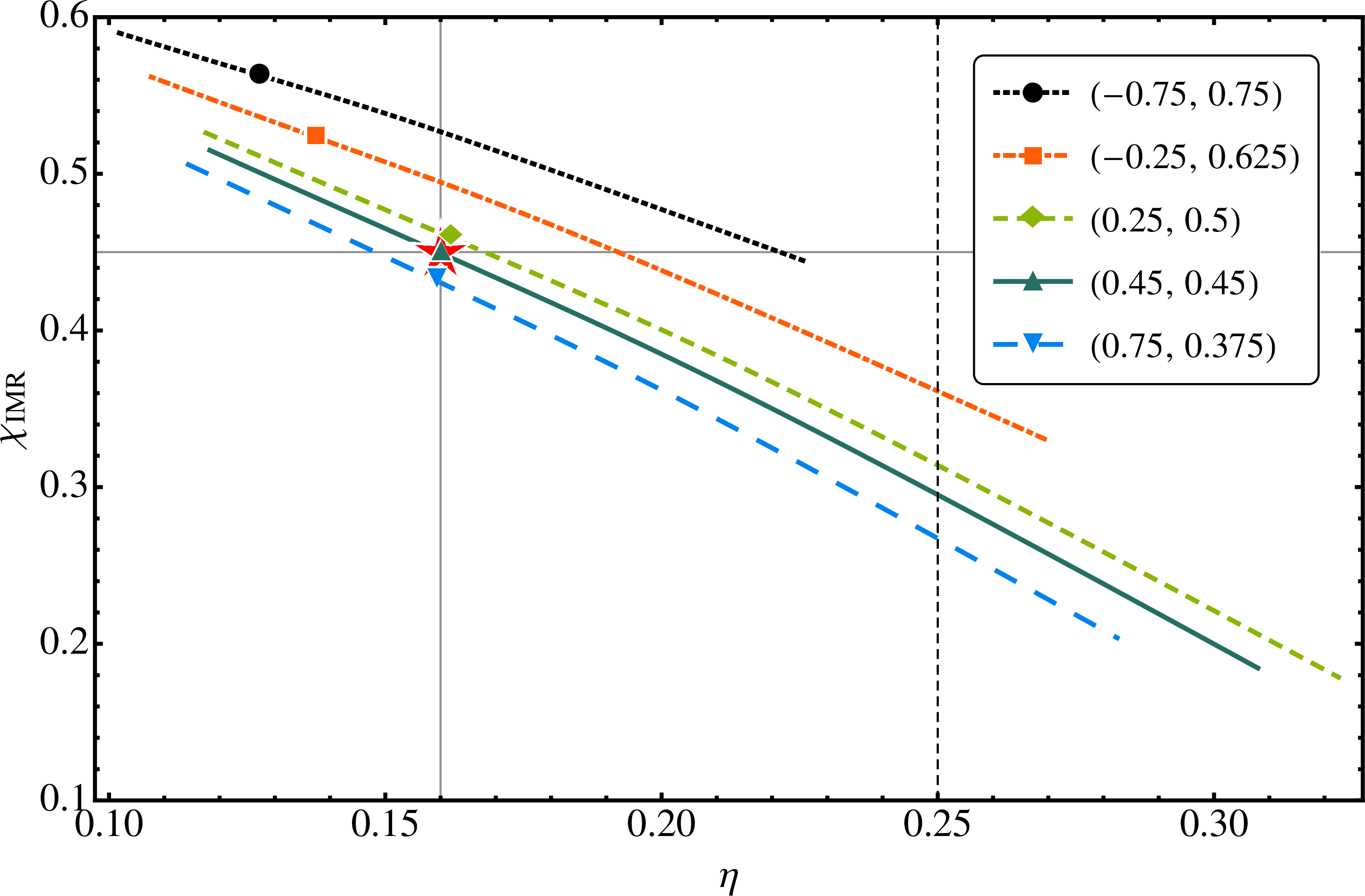}
	\caption{$90\%$ confidence regions represented as lines at $\text{SNR}=50$ and $7M_\odot$ for a single spin PN models using $\chiPN$ (top panel) and $\chiIMR$ (bottom panel) and TaylorF2 signals along $\chiPN=0.401575$. The recovered optimal parameters are denoted by colored symbols, while the true parameters are shown by a red star.}
	\label{fig:plots_PN-model_CRs}
\end{figure}

The non-detection of signals in first-generation detectors suggests that events with SNRs higher than
30 will be rare in second-generation detectors~\cite{Lindblom:2008cm}. This should be borne in mind
when we consider Fig.~\ref{fig:plots_PN-model_CRs}, which shows the 90\% confidence regions 
(see Eq.~\eqref{eq:CR}) for a much higher SNR of 50. 
The confidence regions correspond to a $7M_\odot$ binary, and show results for both single-effective-spin 
PN models. We project the 3-dimensional confidence regions in $(\Mchirp, \eta, \chi)$ onto the symmetric 
mass-ratio $\eta$ and effective spin $\chi$. The optimal parameters are denoted by colored symbols, while 
the true parameters are shown by red stars. At this mass and high SNR the confidence regions are 
very elongated filaments and we choose to depict them by a curve fit through center of the regions. 
The regions shown for the equal-spin configuration $(\chi_1,\chi_2) = (0.45, 0.45)$ (solid dark green) are 
the confidence regions in the proper sense as the signal is exactly represented by the models at this point.
For the waveforms that do not lie in the model subspaces the confidence regions are computed with the model 
waveform that has the best match with the given signal.
The very large uncertainty given by the extent of the confidence regions in Fig.~\ref{fig:plots_PN-model_CRs} 
is due to the (approximate) degeneracy between mass-ratio and 
spin~\cite{Cutler:1994ys,Poisson:1995ef,Baird:2012cu,Hannam:2013uu}. 
One can see from the leading order spin-orbit term that it is possible to mimic the effect of spin by 
modifying the mass-ratio at constant chirp mass.

The uncertainties for the $\chiPN$ model are about $\Delta\chi \sim 0.35$ and $\Delta\eta \sim 0.2$.
For the $\chiIMR$ model the uncertainties are comparable, but a bit larger, roughly $\Delta\chi \sim 0.45$ 
and $\Delta\eta \sim 0.25$. For both models the confidence regions extend into the region of unphysical $\eta > 0.25$.
If we project the confidence regions onto the plane of component masses $(m_1,m_2)$ all configurations 
lie on top of a line of constant chirp mass $\Mchirp \sim 2.33$. For both models the configurations 
range from an equal-mass binary with total mass $5.4 M_\odot$ up to a mass-ratio $q=7$ binary 
with total mass $9 M_\odot$, as opposed to the true parameters $(m_1,m_2) = (1.4, 5.6)$.
These results are consistent with those shown in Ref.~\cite{Hannam:2013uu}, and illustrate the 
point made in that work, that we would not be able to determine if such a source was a binary containing
two black holes, or a black hole and a neutron star. 

We can quantify the additional uncertainty introduced by the single spin approximation with a given 
parameter $\chiPN$ or $\chiIMR$ by comparing the ``spread'' in $\chi$ between the recovered 
parameters with the extent of the model confidence region in the $\chi$ direction. 
For the $\chiPN$-model we find a spread $\Delta\chiPN \sim 0.06$ vs a spread of 
$\Delta\chiIMR \sim 0.13$ for the $\chiIMR$ model. The extent of the equal-spin confidence regions in $\chi$ is 
roughly $0.32$ for $\chiPN$ and $0.34$ for $\chiIMR$. For both models at the chosen mass of 
$7 M_\odot$ the statistical uncertainties dwarf the spread in the biases, even at this high SNR of 50. 
In addition, note that all of the recovered parameters for the $\chiPN$ model are within the statistical 
error bars of the ``true'' parameters. 

These results demonstrate that, while the systematic parameter biases from the single-effective-spin 
models may appear large, they are in fact much smaller than the statistical errors, even at high SNR.
We conclude, then, that the reduced-spin model presented in Ref.~\cite{Ajith:2011ec} is likely to be
sufficient for parameter estimation of low-mass signals from aLIGO and AdV. 
 
		
\section{Results for IMR waveforms} 
\label{sec:IMR-results}

\subsection{Matches between IMR waveforms} 
\label{sec:IMR-matches}

We now consider the family of $\chiIMR = 0.45$ PN-NR hybrid waveforms summarized in 
Tab.~\ref{tab:configs}. 

\begin{figure}[t]
  \centering
    \includegraphics[width=0.5\textwidth]{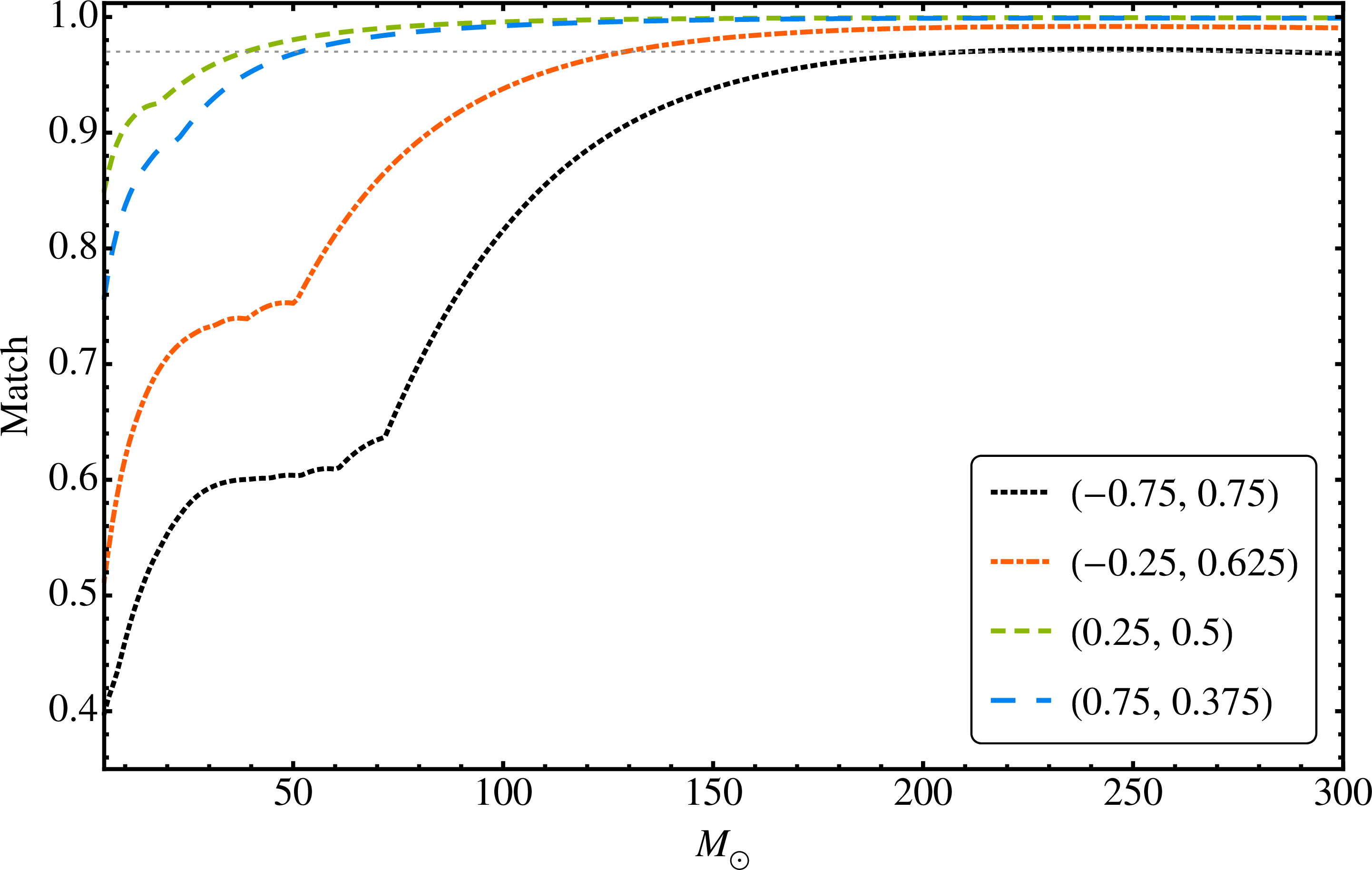}
  \caption{Matches between unequal spin TaylorF2 frequency-domain PN-NR hybrids with the 
  equal spin $(0.45, 0.45)$ hybrid.
}
  \label{fig:plots_TaylorF2_hyb_matches}
\end{figure}

In Fig.~\ref{fig:plots_TaylorF2_hyb_matches} we show matches between the TaylorF2 
frequency-domain hybrid (see Sec.~\ref{sec:hybridization}) of the reference 
$(\chi_1,\chi_2) = (0.45,0.45)$ case, with each of the other configurations listed in 
Tab.~\ref{tab:configs}. 
As we would expect, the further away the individual spins are from the fiducial $(0.45,0.45)$ 
waveform the worse the matches become. 

The degradation of the matches for low masses is expected due to the use of the $\chiIMR$ parameter;
the configurations used in this study lie along a line of $\chiIMR=0.45$. The $(0.45, 0.45)$ configuration 
corresponds to $\chiPN \approx 0.402$. Fig.~\ref{fig:plots_chi_plane} shows how these lines diverge. 
Those configurations for which the spread between these lines is the largest (i.e., those that are the 
farthest away from the fiducial $(0.45, 0.45)$ configuration) are therefore expected to have the worst 
match. In fact, for low masses, the matches between these IMR waveforms are very close to the 
PN matches computed at $7 M_\odot$ in Fig.~\ref{fig:plots_PN-model-1p-matches}, as expected.

Around $50M_\odot$ hybridization artifacts lead to a visible kink in the matches. Note that these are not
artifacts of the hybridization procedure itself, but rather a result of the disagreement between the TaylorF2
and fully general relativistic NR waveforms at the matching frequency; these artifacts could be made 
arbitrarily small if we produced NR waveforms of sufficient length to match to PN at arbitrarily low 
frequencies. Previous studies of NR waveform length requirement (in particular Ref.~\cite{Ohme:2011zm}), 
and the estimates of statistical uncertainties in this study, suggest that the measurement errors 
due to these effects do not have a significant impact on the scientific information that can be extracted 
from aLIGO and AdV GW observations. 

As we go to higher masses the matches improve considerably, which indicates that in the merger 
regime $\chiIMR$ performs well. Beyond $200 M_\odot$ the matches drop off as we move away 
from the fiducial $(0.45, 0.45)$ configuration due to the different final spin and thus different ringdown 
frequency of the remnant BHs.

It is very likely that the matches between waveforms along a line of constant $\chiPN$ would be 
far higher even through the merger and ringdown. Again, our results are consistent with our expectation from PN theory that future phenomenological models should be parametrized by $\chiPN$.

\subsection{Biases and uncertainties against IMRPhenomC} 
\label{sec:IMR-FF-IMRPhenomC}

As with the TaylorF2 reduced-spin PN model in Sec.~\ref{sec:PN-results}, we now wish to study the parameter biases
due to the use of the single-effective-spin $\chiIMR$ in full inspiral-merger-ringdown waveform models. 
For this purpose we will compare our family of constant $\chiIMR = 0.45$ PN-NR hybrids against one of
the current phenomenological IMR models, IMRPhenomC. 
Our goal is complicated by a number of factors. One class of error sources in our analysis is the 
accuracy of the hybrid waveforms: this will depend on the accuracy of the NR waveforms, on the 
frequency at which they are hybridized to PN waveforms, and the accuracy of the PN approximant
that was used. These errors are mitigated by choosing the same PN approximant, TaylorF2, as was
used for the inspiral part of IMRPhenomC. The two other sources of uncertainty (hybridization frequency 
and NR-waveform accuracy), can be quantified, and we will show (in Figs.~\ref{fig:plots_IMRPhenomC_TaylorF2_hyb_biases_errors_NR} and~\ref{fig:plots_IMRPhenomC_TaylorF2_hyb_biases_errors_hybrid}) that they do 
not affect our conclusions. 

A more serious source of error is in the IMRPhenomC model itself. This model was calibrated to 
NR waveforms up to mass ratio 1:4, but \emph{not} spinning-binary waveforms at that mass ratio.
The hybrids that we compare with the model are therefore at the very edge of the region of parameter
space over which the model was calibrated. In addition, IMRPhenomC (and all current phenomenological 
models in general) were designed with detection in mind, and not as a tool for parameter estimation. 
There are certainly errors in how well IMRPhenomC represents the hybrid waveforms from which it 
was built, i.e. \emph{modeling errors}, as well as hybridization artifacts due to the waveform length, and
errors in the NR simulations that were used. 
From these combined error sources we expect a bias in the parameters that we estimate using this 
model, which will be nontrivially combined with the errors due to the use of the single-effective-spin
approximation, which are the biases we wish to measure in this study. The other error sources can
be reduced in future phenomenological models, while the bias due to the single-effective-spin approximation
will be inherent in all such models. 

Despite these complications, we are able to draw a number of important conclusions from our results, 
which we discuss in this section. 

\begin{figure}[t]
  \centering
    \hspace*{-2mm}\includegraphics[width=0.5\textwidth]{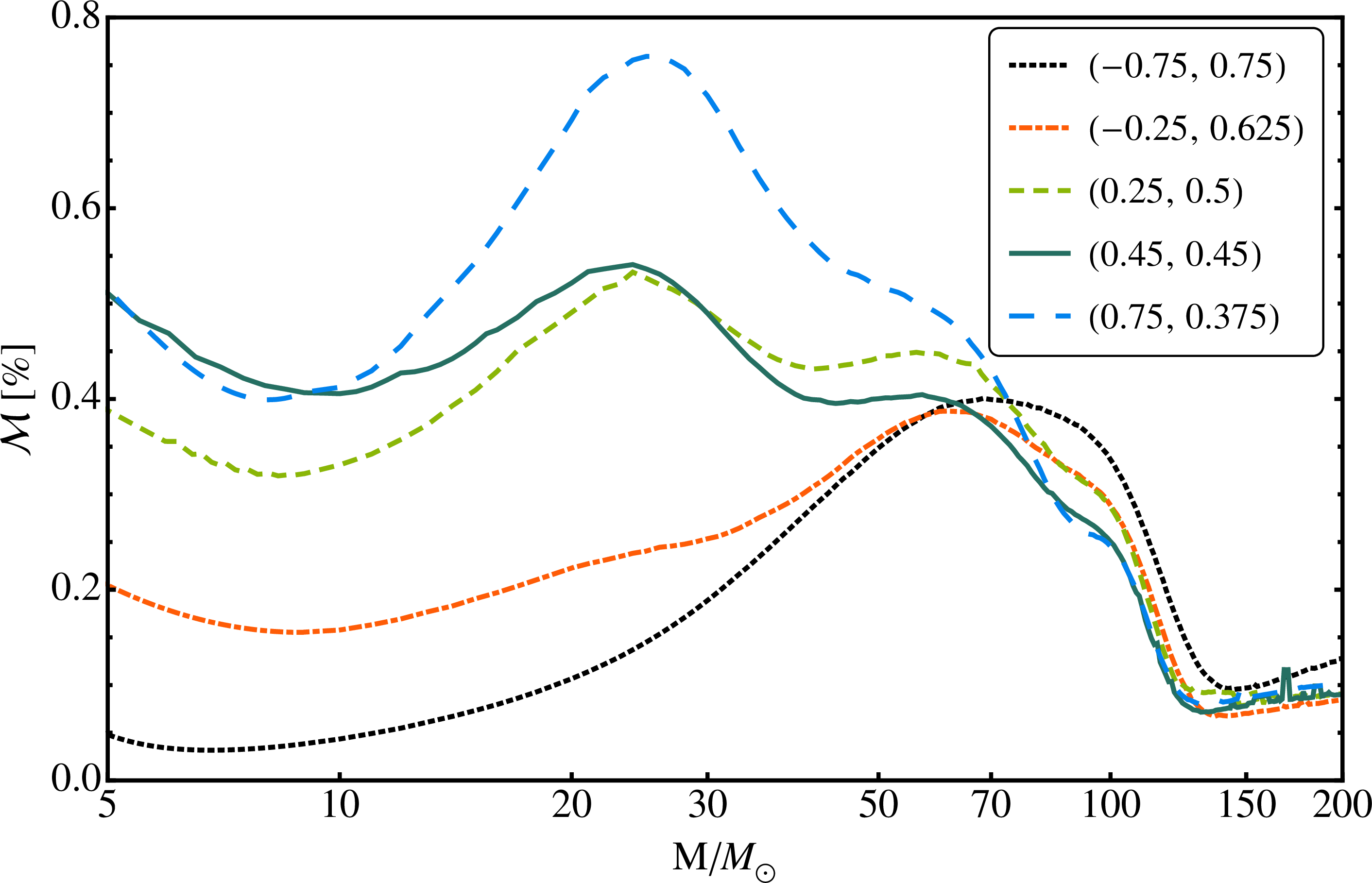}
  \caption{Fully optimized mismatches between IMRPhenomC and TaylorF2 PN-NR hybrids.}
  \label{fig:plots_IMRPhenomC_TaylorF2_hyb_FFs}
\end{figure}

The first point to emphasize is that, despite all of the shortcomings outlined above, the IMRPhenomC 
model achieves its main purpose as a search template family. 
We show fully optimized mismatches with IMRPhenomC in Fig.~\ref{fig:plots_IMRPhenomC_TaylorF2_hyb_FFs}. 
All of them are below $1\%$. This confirms the effectualness of IMRPhenomC in this region of the 
parameter space, and the suitability of the model for aligned-spin GW searches. 
Let us contrast these results of IMRPhenomC with its predecessor IMRPheomB: Due to its simpler PN part 
IMRPhenomB is only effectual for masses $>20 M_\odot$ for the configurations considered here. 

\begin{figure}[t]
  \centering
    \hspace*{-2mm}\includegraphics[width=0.5\textwidth]{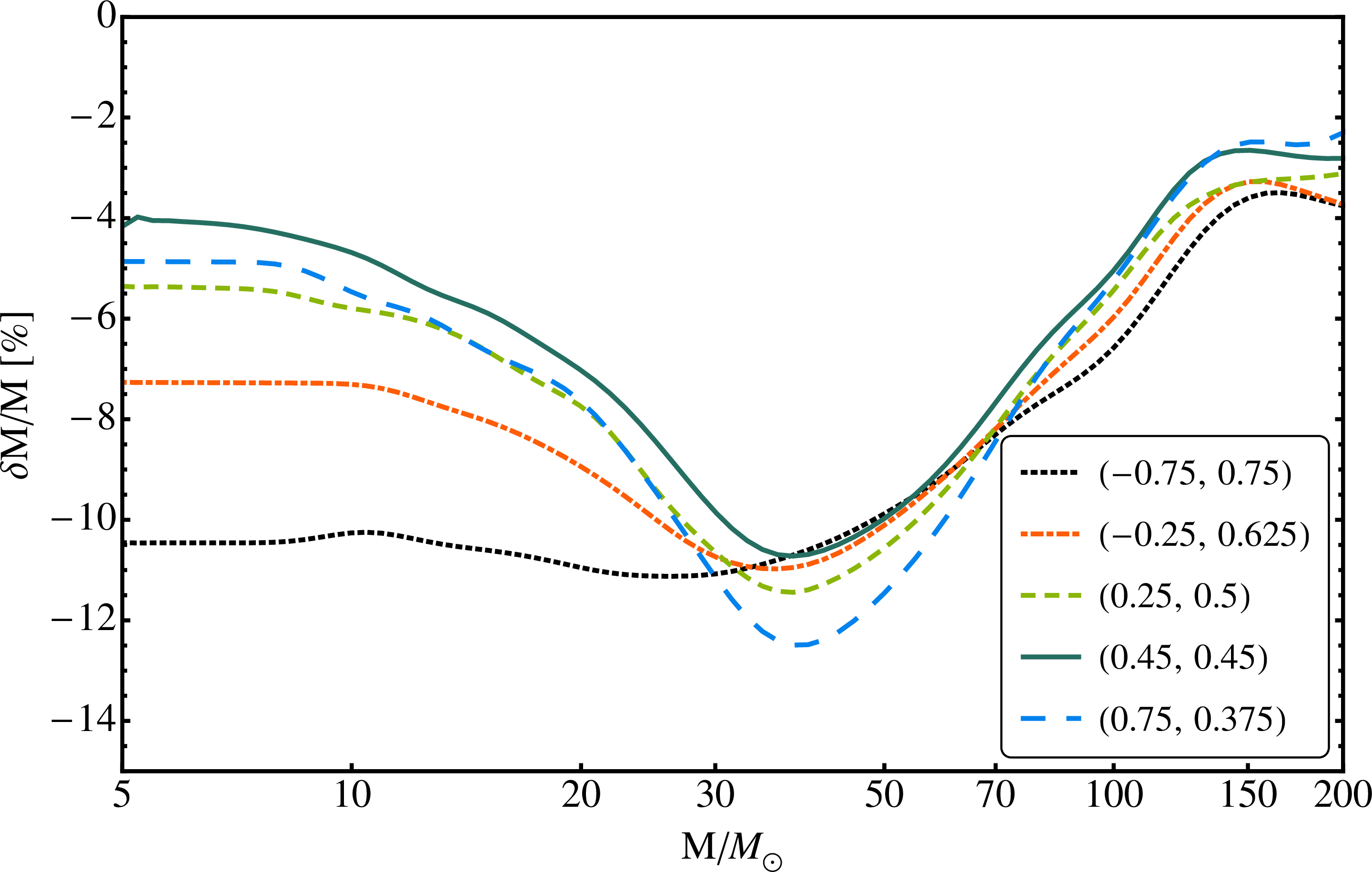}
  \caption{Biases in the binary mass for IMRPhenomC and TaylorF2 PN-NR hybrid waveforms.}
  \label{fig:plots_IMRPhenomC_TaylorF2_hyb_mass_bias}
\end{figure}

Biases of the total binary mass, as a function of the total mass of the signal, are shown in 
Fig.~\ref{fig:plots_IMRPhenomC_TaylorF2_hyb_mass_bias}. This figure illustrates
well the complications that were discussed above in our comparisons with IMRPhenomC. The reference
waveform has spins $(0.45, 0.45)$, and should be identical to the $\chiIMR = 0.45$ waveform in 
the model. The bias in the total mass for this case is therefore most likely due to modeling artifacts in 
IMRPhenomC. If we are to assess the bias due only to the use of the single-effective-spin approximation, 
then we must look at the \emph{spread} of the parameters away from the (solid) $(0.45, 0.45)$ line
in the figure. We then see that the spread in the total mass can be as high as $\sim$5\% at low masses 
(due in most part, once again, to the parametrization by $\chiIMR$ instead of $\chiPN$) 
and no more than $\sim$2\% at intermediate masses. At high masses the spread in the total mass
is around 1\%. 

\begin{figure}[t]
  \centering
    \hspace*{-2mm}\includegraphics[width=0.515\textwidth]{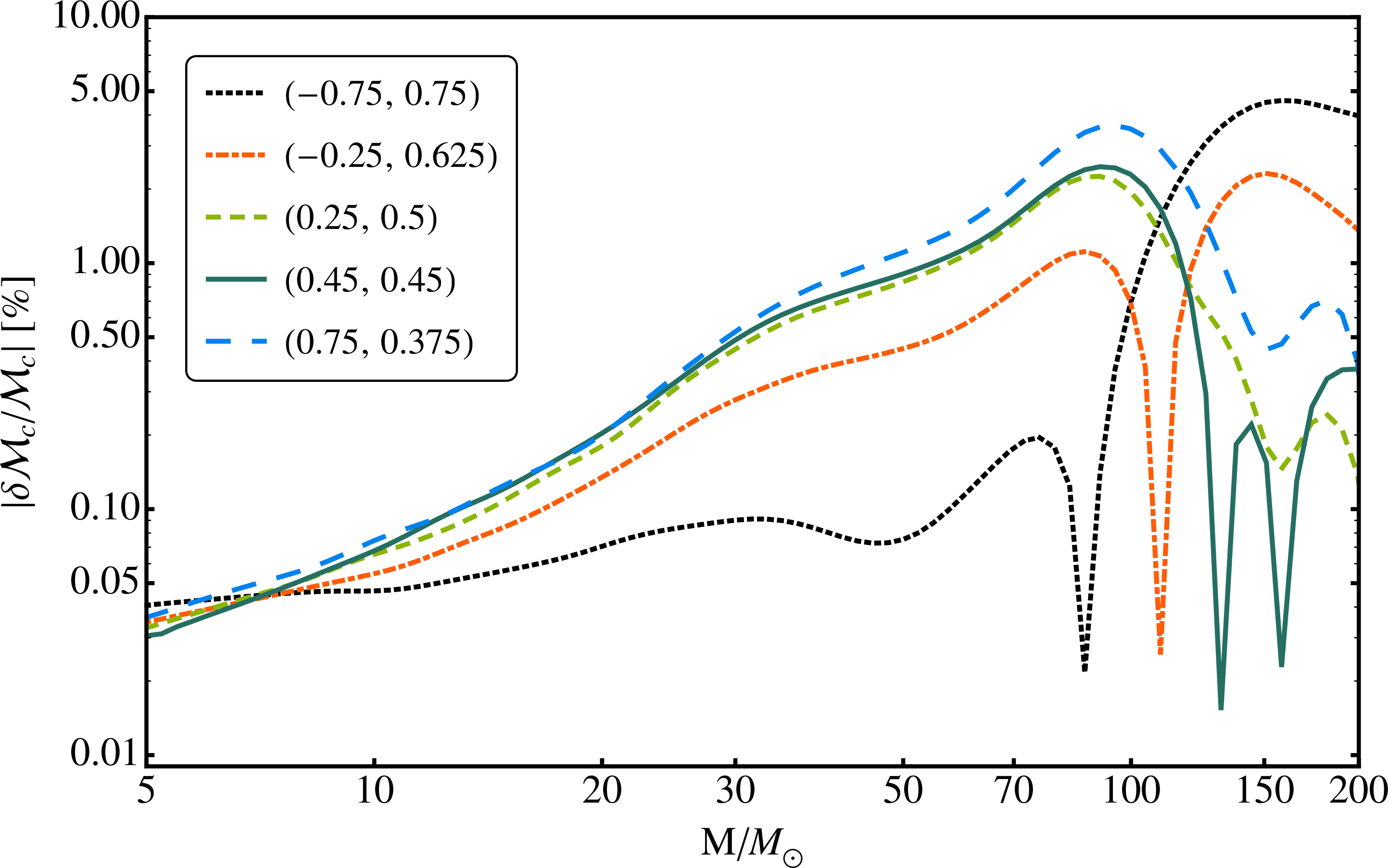}
  \caption{Biases in the chirp mass $\Mchirp$ for IMRPhenomC and TaylorF2 PN-NR hybrid waveforms.}
  \label{fig:plots_IMRPhenomC_TaylorF2_hyb_chirp_mass_bias}
\end{figure}

The bias in the chirp mass is shown in Fig.~\ref{fig:plots_IMRPhenomC_TaylorF2_hyb_chirp_mass_bias}. 
For inspiral signals, we expect to be able to measure the
chirp mass extremely accurately, because it is the leading-order PN contribution; that
is the motivation for the definition of the chirp mass. And indeed the bias in the chip 
mass is below 0.1\% for masses below 10\,$M_\odot$. However, the chirp mass has little significance 
during merger and ringdown, and the chirp-mass bias increases for higher-mass binaries; 
the fractional error in the chirp mass is 
comparable to (and in some cases larger than) that in the total mass above 50\,$M_\odot$. 
Note that the dips in the curves in the figure are due to the use of a logarithmic scale; these are
points where $\delta \mathcal{M}/\mathcal{M}$ changes sign.

\begin{figure}[htbp]
  \centering
    \hspace*{-1.8mm}\includegraphics[width=0.5\textwidth]{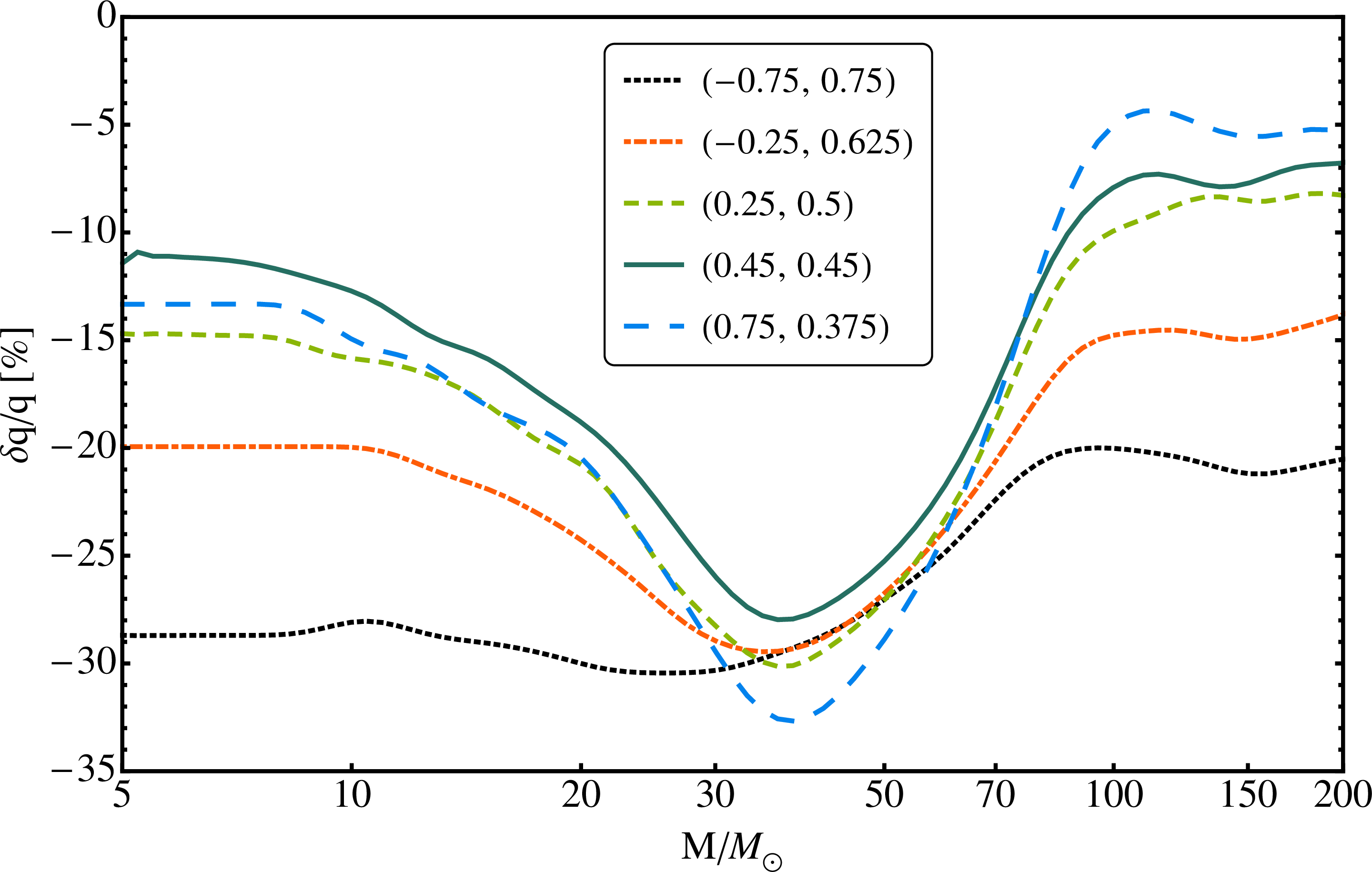}
    \includegraphics[trim=0mm 0mm 0mm -5mm,clip,width=0.49\textwidth]{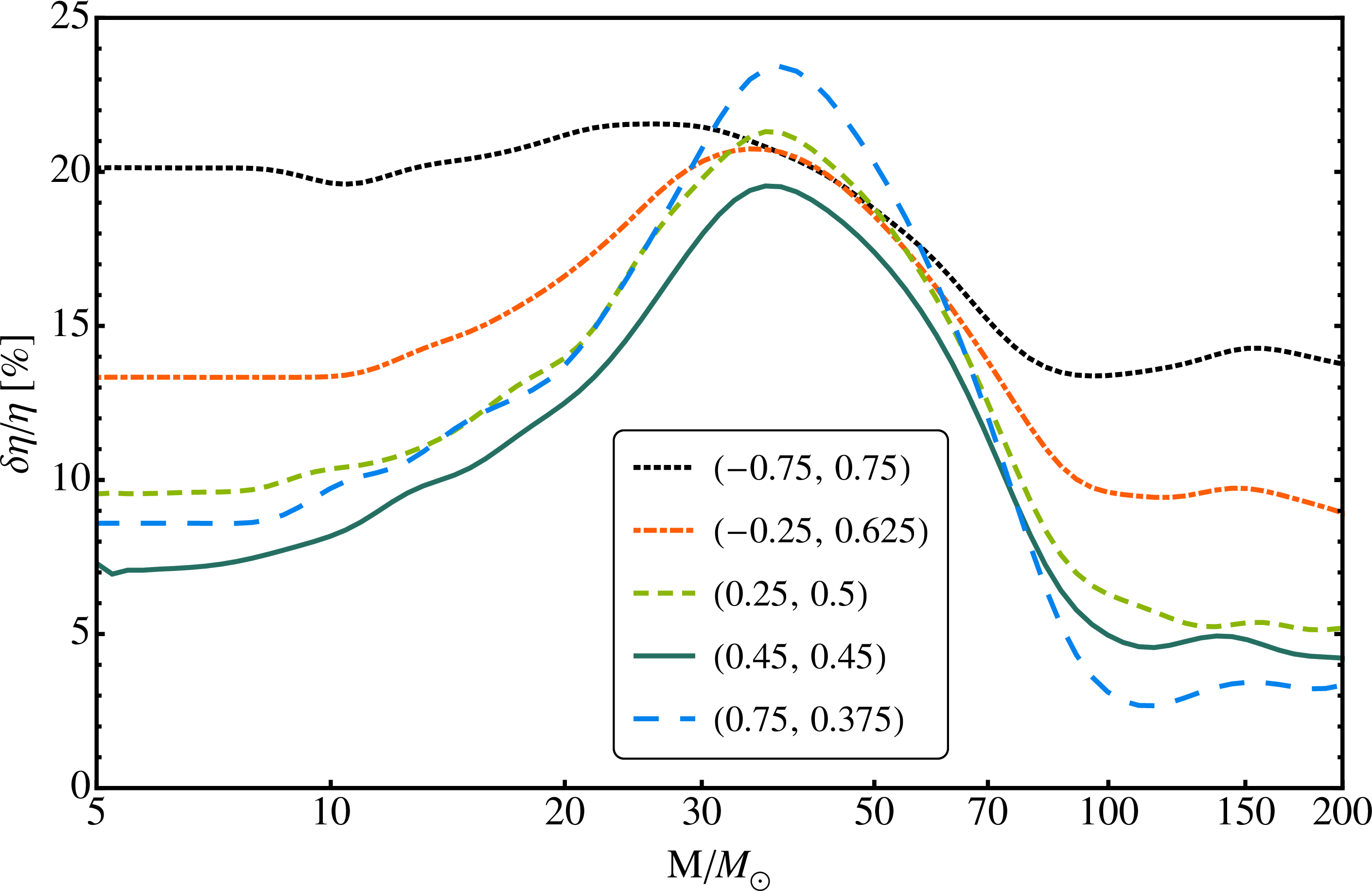}
  \caption{Biases in the mass-ratio and the symmetric mass-ratio for IMRPhenomC and TaylorF2 PN-NR hybrid waveforms.}
  \label{fig:plots_IMRPhenomC_TaylorF2_hyb_mass_ratio_bias}
\end{figure}

The biases in mass ratio and the symmetric mass-ratio are shown in 
Fig.~\ref{fig:plots_IMRPhenomC_TaylorF2_hyb_mass_ratio_bias}. Once again, to interpret these figures in
terms of the bias due to the single-effective-spin approximation, we should consider the spread of values
around the $(0.45, 0.45)$ lines. We see for the mass ratio, the spread in values is around 20\% for high-mass
systems. (At low masses, the results are again exaggerated by the use of the $\chiIMR$ parameter.) 

\begin{figure}[htbp]
  \centering
    \hspace*{-0.5mm}\includegraphics[width=0.503\textwidth]{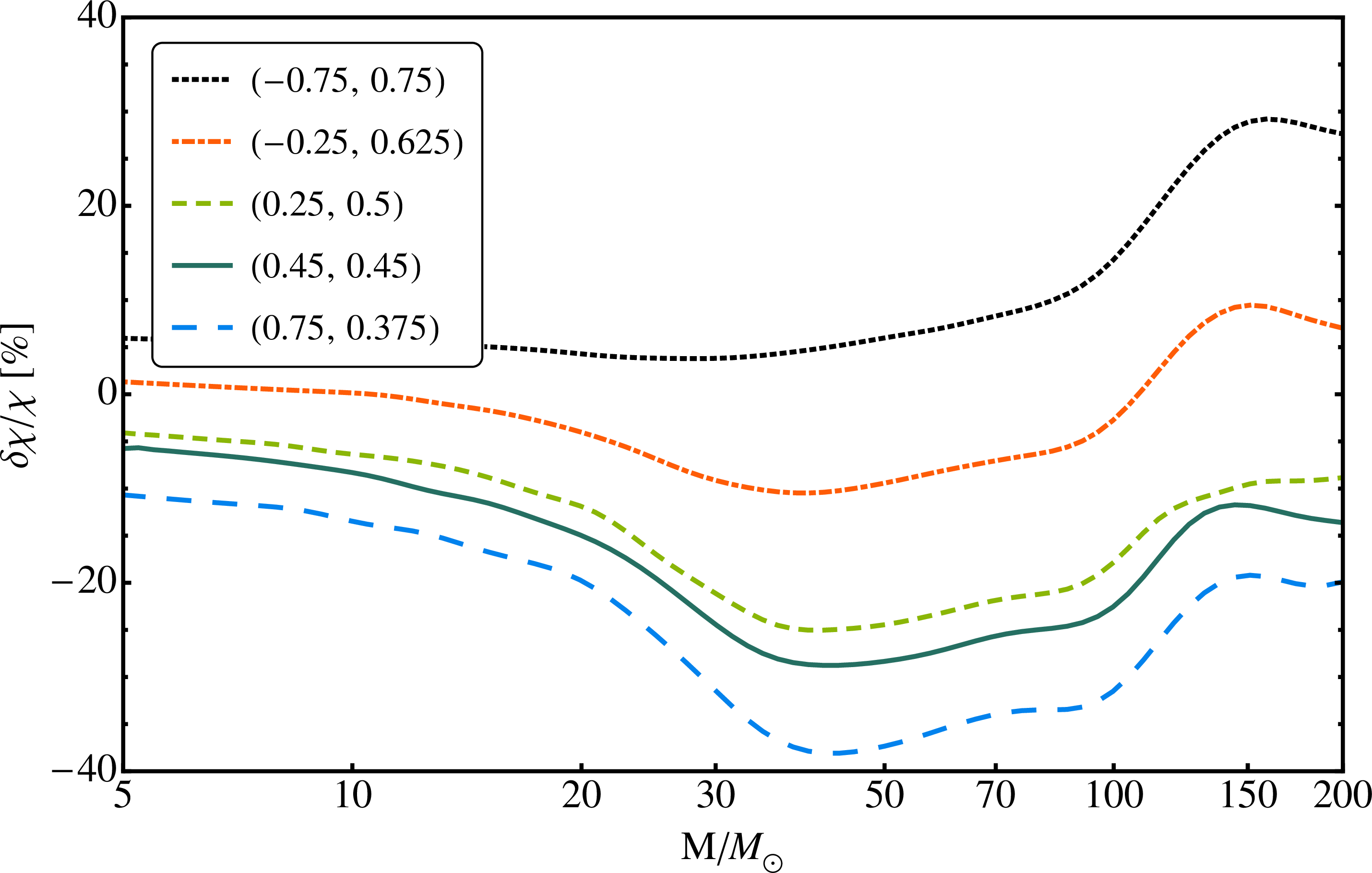}
    \includegraphics[width=0.5\textwidth]{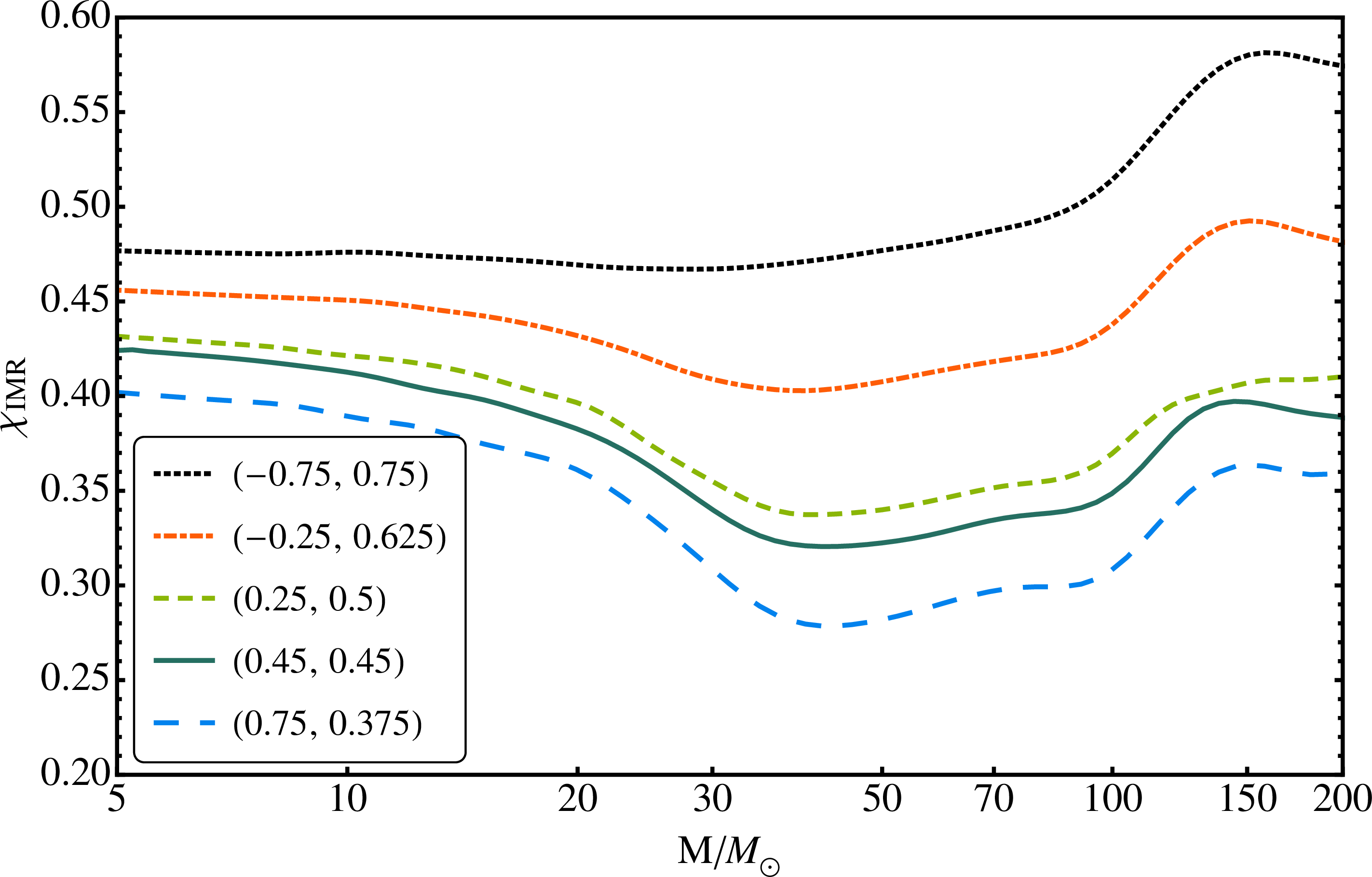}
  \caption{Biases in the effective spin $\chiIMR$ for IMRPhenomC and TaylorF2 PN-NR hybrid waveforms (top) and recovered
 effective spin $\chiIMR$ (bottom). At SNR 10 the statistical uncertainty from the $90\%$ confidence region for masses $20, 50, 100 M_\odot$ is roughly $0.15$, $0.2$ and $0.4$, respectively.
}
  \label{fig:plots_IMRPhenomC_TaylorF2_hyb_spin_bias}
\end{figure}

The biases in the spin parameter and the recovered spin values themselves are shown in 
Fig.~\ref{fig:plots_IMRPhenomC_TaylorF2_hyb_spin_bias}.
At low masses it makes sense to compare the results with the findings of Sec.~\ref{sec:PN-results} where 
we studied single spin PN models. The PN model using $\chiIMR$ is the relevant one to compare with. 
The biases are in general consistent between our PN and IMR studies: very low in $\Mchirp$, of 
similar magnitude in $\eta$ and $\chi$. The spread in the $\chi$ biases is a little smaller, about $15\%$ 
for the IMR results, as opposed to $25\%$ for PN. The difference in the PN and IMR confidence regions,
even at low masses, was also noted in Ref.~\cite{Hannam:2013uu}, and will be studied further in 
future work.

\begin{table}[t] 
  {\footnotesize
		\newcolumntype{.}{D{.}{.}{-1}}
	  \begin{tabular}{lc....}
	  	\hline\hline
	    \multicolumn{1}{c}{Case $(\chi_1,\chi_2)$} & 
			\multicolumn{1}{c}{Mismatch $\MM$} & 
			\multicolumn{1}{c}{$\Delta \Mchirp/\Mchirp [\%]$} &
	    \multicolumn{1}{c}{$\Delta \eta /\eta [\%]$}  & 
			\multicolumn{1}{c}{$\Delta \chi /\chi [\%]$} &
			\multicolumn{1}{c}{recovered $\chiIMR$} \\
	    \hline
		 (-0.75,0.75) 	& $4\times 10^{-3}$ & -0.1 	& 17.65 & 5.15   & 0.473\\
		 (-0.25,0.625) 	& $4\times 10^{-3}$ & -0.34 & 15.35 & -6.67  & 0.42\\
		 (0.25,0.5) 		& $4\times 10^{-3}$ & -0.88 & 17.51 & -28.93 & 0.32\\
		 (0.45,0.45) 		& $4\times 10^{-3}$ & -0.89 & 15.56 & -32.07 & 0.306\\
		 (0.75,0.375) 	& $5\times 10^{-3}$ & -1.06 & 17.83 & -44.06 & 0.252\\
	    \hline\hline
	   \end{tabular}
  }
  \caption{Fully optimized mismatches and biases between IMRPhenomC and IMR signals (frequency domain TaylorF2 hybrids) along $\chiIMR=0.45$ computed using local minimization in the center of the confidence region for $50M_\odot$.}
    \label{tab:IMRPhenomC_FF_biases_50MS}
\end{table}

\begin{figure}[t]
  \centering
		\includegraphics[trim=0mm 0mm 0mm -8mm,clip,width=0.5\textwidth]{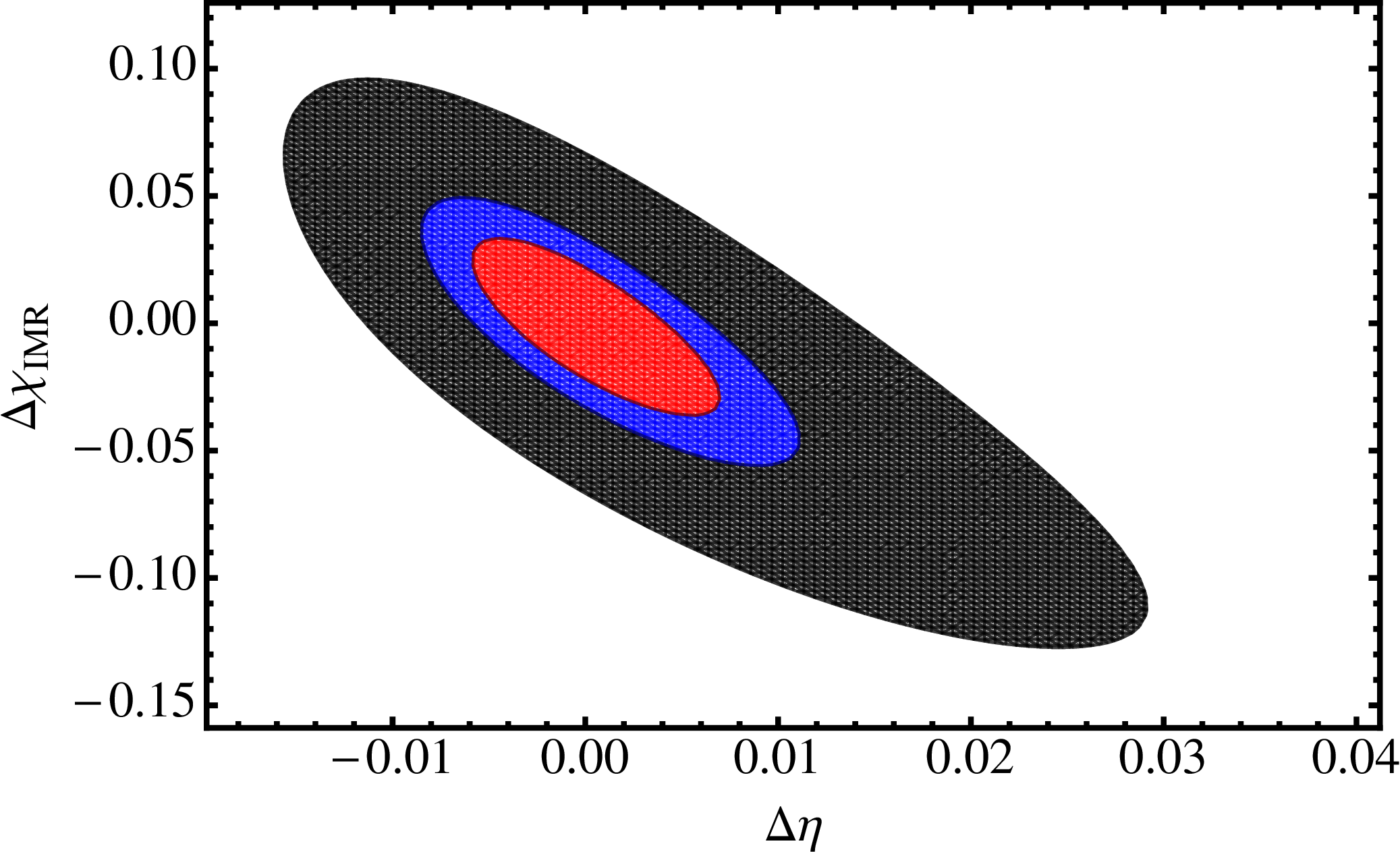}
  \caption{$90\%$ confidence regions at $\text{SNR}=10,20,30$ (black, blue, red) and $50M_\odot$ for IMRPhenomC with IMRPhenomC $(\eta=0.16, \, \chi=0.45)$.
}
  \label{fig:plots_IMRPhenomC_self_CRs_50M_SNR20}
\end{figure}

As in the PN study in Sec.~\ref{sec:PN-results}, the spread in biases needs to be put into context 
with the statistical uncertainty in the parameter measurements. For comparison,
Fig.~\ref{fig:plots_IMRPhenomC_self_CRs_50M_SNR20} shows the $90\%$ confidence regions (see Eq.~\eqref{eq:CR}) 
for a 50\,$M_\odot$ binary and table~\ref{tab:IMRPhenomC_FF_biases_50MS} summarizes the biases at this mass. 
While the extent of the confidence regions is much more confined than was the case for the PN models, the regions are still elongated in a diagonal direction in $\eta$ and $\chi$ and illustrate the degeneracy between mass-ratio and spin~\cite{Baird:2012cu}.
We see that the uncertainty in the masses is in general larger 
than the spread in the parameter biases. For example, the spread in $\eta$ values at 50\,$M_\odot$
was $\sim$5\%, while the statistical uncertainty in the mass ratio at SNR 30 is $\sim$10\% (see Fig.~\ref{fig:plots_IMRPhenomC_self_CRs_50M_SNR20}). We can
conclude, then, that the single-effective-spin approximation does not adversely affect estimation of 
the black-hole masses at likely advanced-detector SNRs. 

\begin{table}[t] 
  {\footnotesize
	  \begin{tabular}{lcccc}
	  	\hline\hline
			Mass  															& $20 M_\odot$ 	& $50 M_\odot$ 	& $100 M_\odot$ \\
	    \hline
			Spread in $\chi$ biases 						&	0.11					&	 0.22 				& 0.2  \\
			Extent of CR in $\chi$ SNR 10 			& 0.16					&  0.22 				& 0.41 \\
			Extent of CR in $\chi$ SNR 20 			& 0.07					&  0.1  				& 0.2  \\
			Comparable SNR 											& 13						&  10   				& 20   \\         
	    \hline\hline
	   \end{tabular}
  }
  \caption{We compare the spread in the $\chi$ biases for the 5 hybrids with IMRPhenomC against the extent of the
  IMRPhenomC confidence region in the $\chi$ direction at SNR 10 and 20 for $M=20,50,100 M_\odot$. We also show at 
	which SNR these numbers become comparable. 
}
  \label{tab:chi_spread_CR_comparison}
\end{table}

The situation is quite different for the spin parameter. At 50\,$M_\odot$, the recovered spin parameter
has a spread of $\Delta\chiIMR \approx 0.2$, while the statistical uncertainty in $\chiIMR$ becomes 
comparable at SNR 10. For higher SNRs the spread in the spin bias dominates.

\begin{figure}[t]
  \centering
    \includegraphics[trim=8mm 0mm 0mm 1.1cm,width=0.48\textwidth]{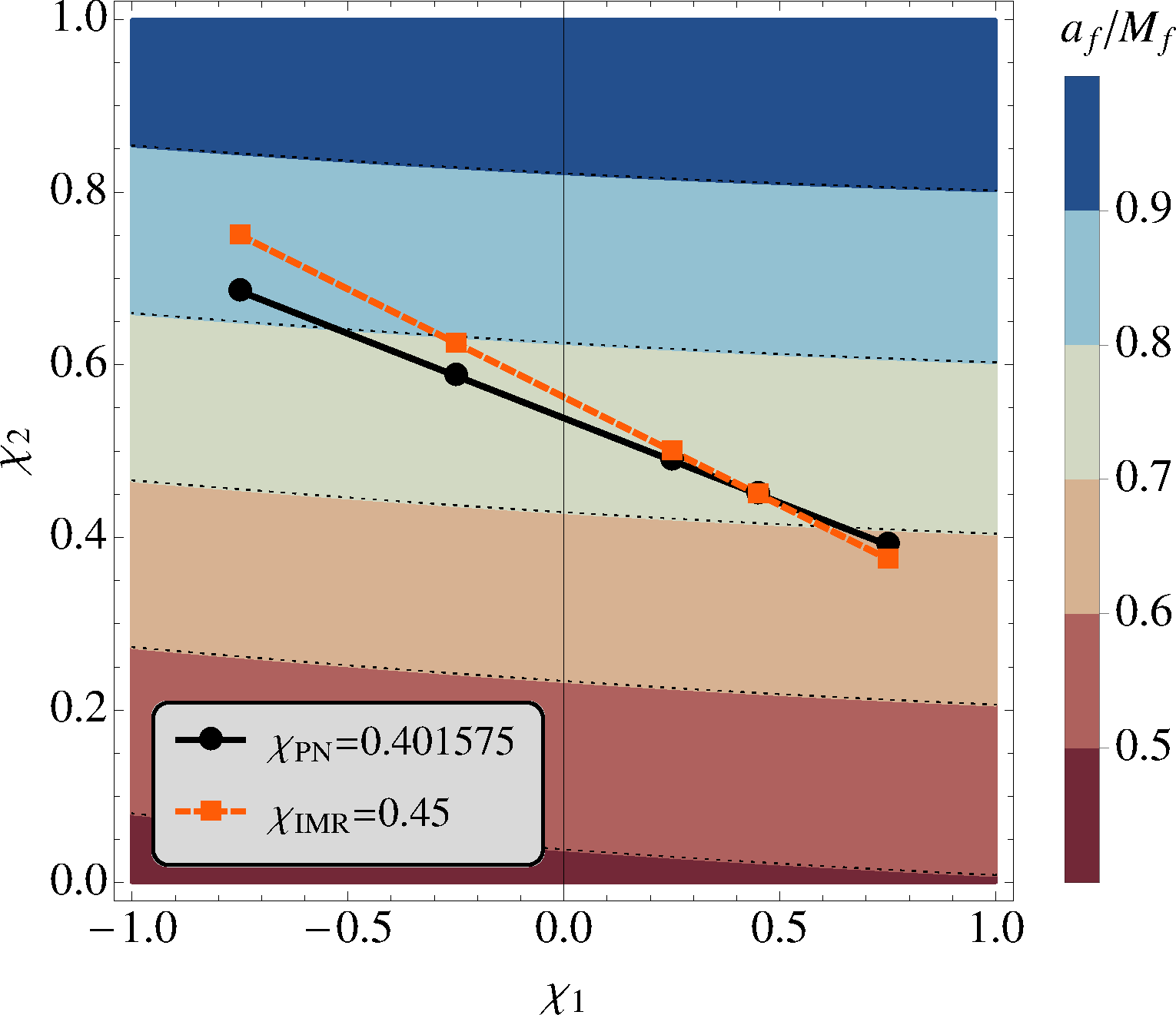}
  \caption{Contours of constant final spin $a_f/M_f$ compared to lines of $\chi=const$ as shown in 
  Fig.~\ref{fig:plots_chi_plane}.}
  \label{fig:plots_final_spin_vs_const_chi}
\end{figure}

We focus on the bias in $\chiIMR$ in Tab.~\ref{tab:chi_spread_CR_comparison}. We concentrate on 
the spread of the biases and therefore ignore the offset in the average bias of the IMR configurations 
from the true value, due to the model not being faithful in this region of the parameter space. 
The spread of recovered parameter values gives us an indication of the additional uncertainties 
introduced by the single-effective-spin approximation.
We find that the single spin approximation is valid for $\chiIMR$ to SNR 10 for masses $7,20,50 M_\odot$ 
and up to SNR 20 for $M>100 M_\odot$. Results for IMRPhenomB lead to comparable conclusions.

We have noted in Sec.~\ref{sec:IMR-matches} that the waveform from the ringdown of the final black hole will 
be characterized by the final spin, and not by either $\chiPN$ or $\chiIMR$. We illustrate this 
point in Fig.~\ref{fig:plots_final_spin_vs_const_chi}, where we overlay the curves of constant $\chiPN = 0.401575$ and 
$\chiIMR = 0.45$ on the contours of constant \emph{final spin}, which can be predicted by a number 
of formulas in the literature~\cite{Tichy:2008du,Barausse:2009uz,Rezzolla:2007rz,Buonanno:2007sv,Lousto:2009mf}.
The results of the various final-spin formulas agree to 
within a few percent with our numerical results for the final spins (see Tab.~\ref{tab:configs}) with the 
largest disagreement at the (-0.75,0.75) configuration.
We see that configurations with the same value of
either spin parameter during inspiral can lead to a black hole with a wide range of final spins, 
depending on the individual spins of the progenitor black holes. (In Tab.~\ref{tab:configs} we see
that final spins for our family of $\chiIMR = 0.45$ numerical simulations ranges from 0.68 up to
0.84.) 

\begin{figure}[t]
  \centering
    \includegraphics[width=0.5\textwidth]{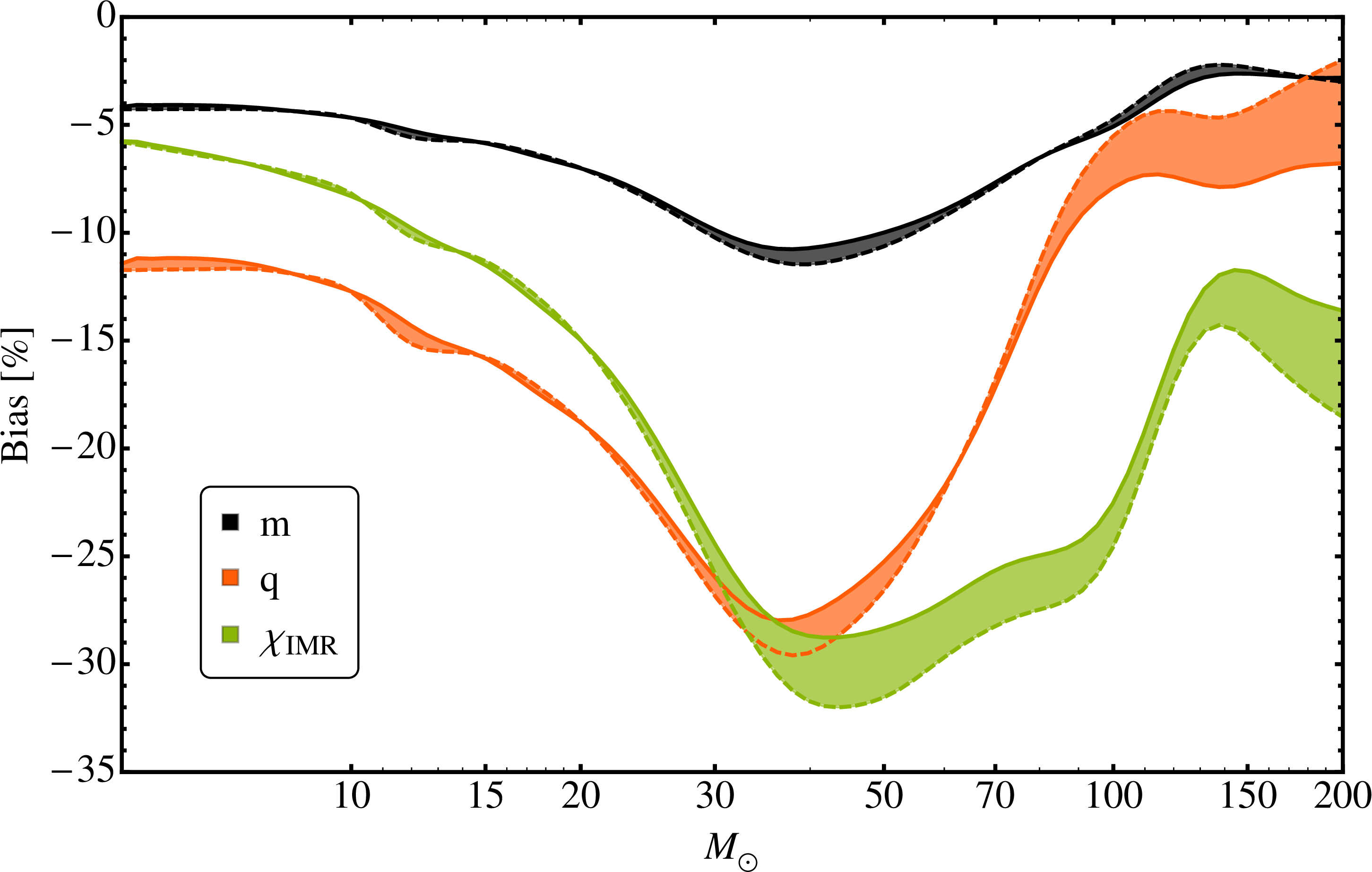}
  \caption{Biases for the $(0.45, 0.45)$ configuration using a numerical resolution of $N=80$ (solid) vs $N=64$ (dashed) gridpoints.
}
  \label{fig:plots_IMRPhenomC_TaylorF2_hyb_biases_errors_NR}
\end{figure}

\begin{figure}[t]
  \centering
		\includegraphics[width=0.5\textwidth]{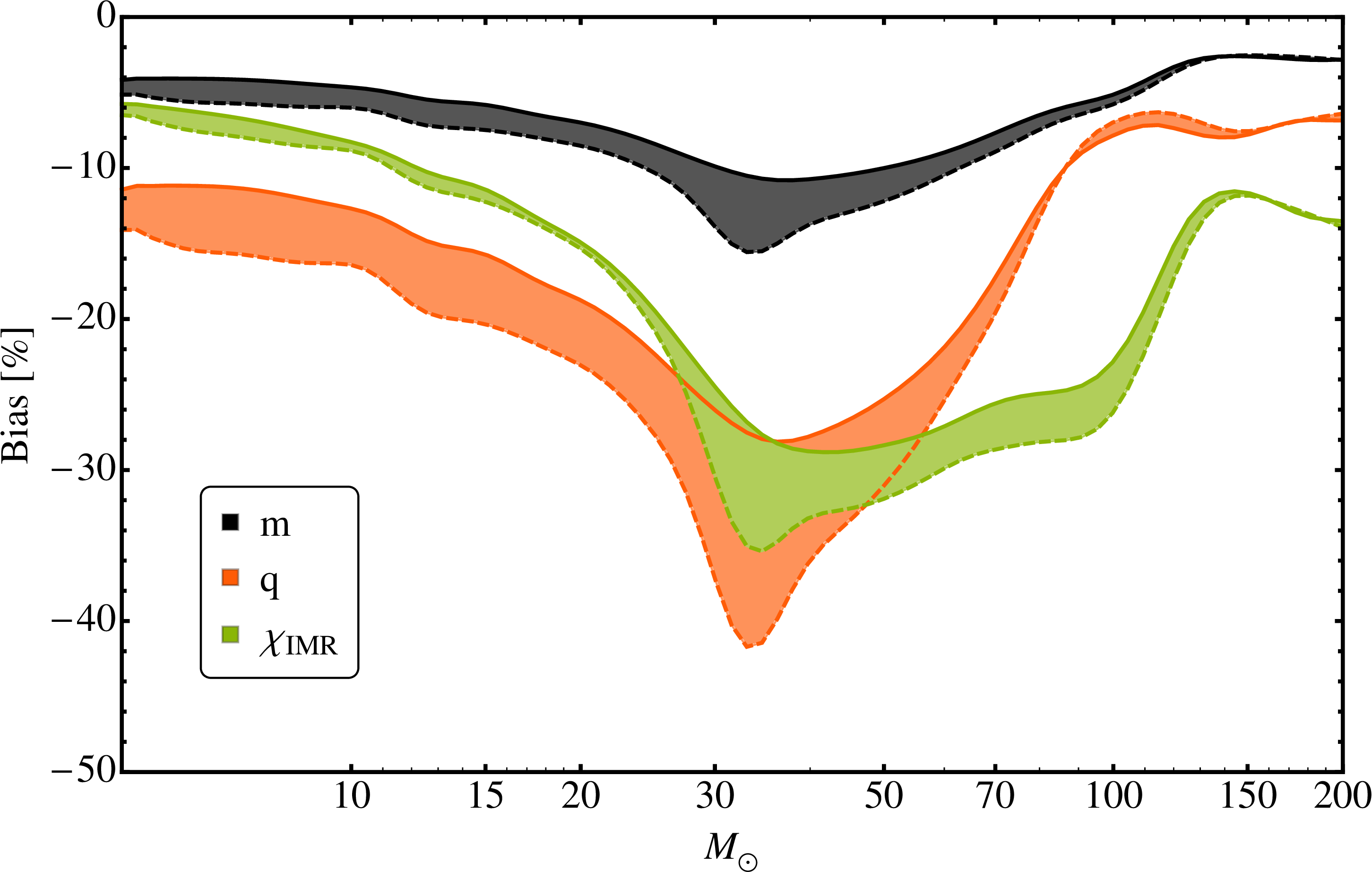}
  \caption{Biases for the $(0.45, 0.45)$ configuration using TaylorF2-hybrids with matching frequencies of $M\omega_m = 0.07$ (solid) vs $M\omega_m = 0.08$ (dashed).
}
  \label{fig:plots_IMRPhenomC_TaylorF2_hyb_biases_errors_hybrid}
\end{figure}

Finally, we verify that our results are not qualitatively changed by errors in the numerical-relativity waveforms
or artifacts due to the choice of hybridization frequency. 
Fig.~\ref{fig:plots_IMRPhenomC_TaylorF2_hyb_biases_errors_NR} compares the variation in biases caused by  
changing the resolution of the numerical waveform used to construct the hybrid, while Fig.~\ref{fig:plots_IMRPhenomC_TaylorF2_hyb_biases_errors_hybrid} shows the effect of changing the 
hybridization frequency. Changing the NR resolution only manifests itself at higher masses 
(both hybrids have been constructed at comparable matching frequencies $M\omega \sim 0.07$), 
whereas changing the hybrid parameter affects predominantly low to medium masses. While this is not 
an in-depth error analysis, it gives an indication of how sensitive our results are to these two sources of 
errors that we can control. It is clear from the figures that neither of these error sources appears to be a serious issue in most cases, but does warrant further study in the future.


\section{Conclusions} 
\label{sec:conclusion}

Several non-precessing-binary waveform models make use of the observation that the 
effects of the black-hole spins on the inspiral rate can be approximated by a single effective 
spin parameter, either $\chiPN$ in the case of a reduced-spin inspiral model~\cite{Ajith:2011ec}, 
or $\chiIMR$ for phenomenological inspiral-merger-ringdown models~\cite{Ajith:2009bn,Santamaria:2010yb}. 
(The two parameters are defined in Eqns.~\eqref{eq:chi_PNdef} and~\eqref{eq:chi_IMRdef}.)
These models were developed primarily for use in template banks
for gravitational-wave searches, but they have also been used for parameter 
estimation~\cite{Aasi:2013jjl}. 
We have investigated the systematic bias in parameter measurement due to the use of these 
single-effective-spin approximations. 

Our primary goal has been to explore the parameter bias from phenomenological waveform models,
and for this purpose we focussed on a family of numerical-relativity simulations of binaries with mass
ratio 1:4, and all with $\chiIMR = 0.45$ (but with different values of individual spins). 
Fig.~\ref{fig:plots_TaylorF2_hyb_matches} shows that the noise-weighted
inner product (match) between these waveforms is in general larger than 0.97 for masses greater 
than 200\,$M_\odot$, but the match degrades significantly at lower masses.  
We note that if we had parameterized these waveforms instead by a constant value of $\chiPN$, then their
matches would be much better at low masses, since we know that the waveforms are partially 
degenerate in $\chiPN$ in the low-frequency (post-Newtonian) regime. 

\begin{figure}[htbp]
  \centering 
  \begin{overpic}[width=0.49\textwidth,scale=1.0]{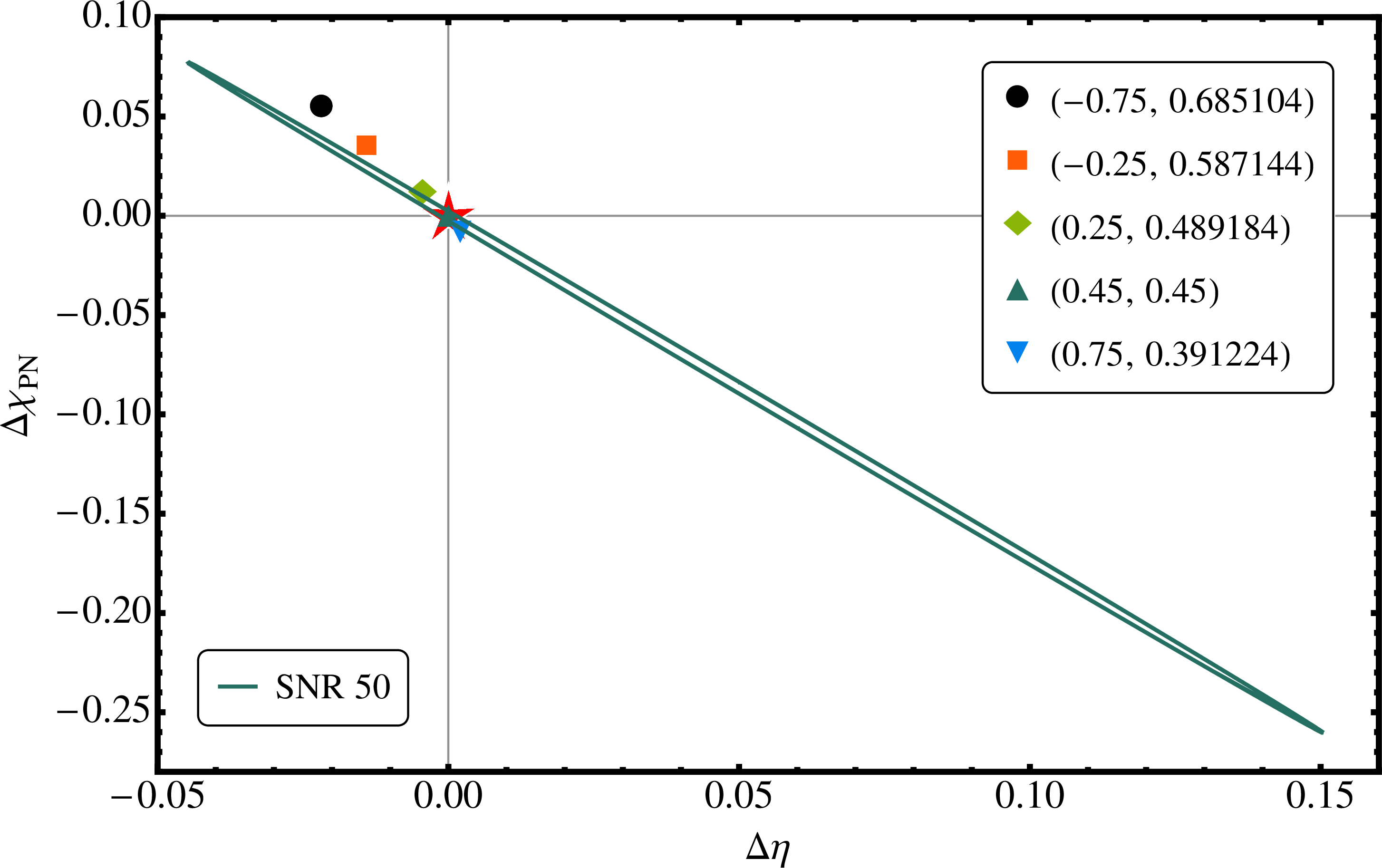}
      \put(42,44){Inspiral}
   \end{overpic}
   \begin{overpic}[width=0.5\textwidth,scale=1.0]{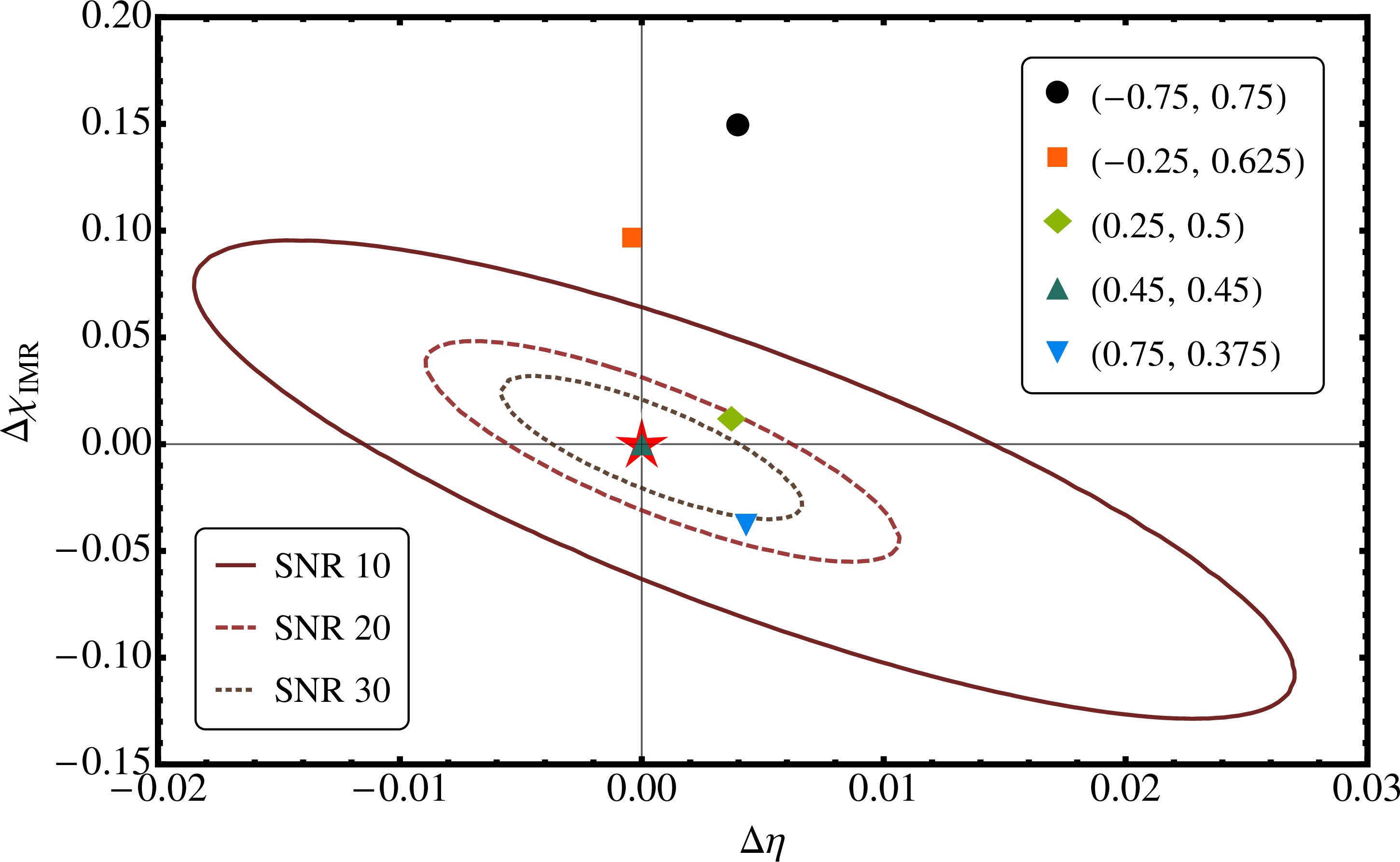}
       \put(48,65){IMR}
    \end{overpic}
  \caption{
These plots compare the systematic biases in measuring the effective spin parameter $\chi$ and the 
symmetric mass ratio $\eta$ with the corresponding statistical uncertainties in the case of the 
\emph{reduced-spin} PN model (top) and the phenomenological IMR model IMRPhenomC (bottom). 
In each panel, the star corresponds to the actual value of parameters ($\eta=0.16, \chiPN = 0.401575$ 
in the top panel and $\eta=0.16, \chiIMR=0.45$ in the bottom panel) while the other markers correspond 
to the parameters of the best-fit single-spin model. The ellipses correspond to the statistical uncertainty 
in measuring the parameters (90\% confidence region) at different SNRs. In the bottom panel, the 
recovered parameter values have been shifted so that they coincide with the true parameters for the 
equal-spin configuration.
}
  \label{fig:CRs_and_Biases_summary}
\end{figure}

In assessing systematic parameter biases, we first consider the inspiral regime, and study 
two related families of post-Newtonian waveforms, one with $\chiPN = 0.45$ and the other 
$\chiIMR = 0.45$ (again, with varying individual spins). Here we find that, for signal-to-noise ratios below 50 
(and we expect most observations in aLIGO and AdV to be below 30), the statistical uncertainty in the
measurement of both the masses and the spin ($\chiPN$) is significantly larger than the parameter bias
incurred by using the single-effective-spin-approximation model to estimate the parameters. This result is
discussed in detail in Sec.~\ref{sec:PN-results}, and summarized here in the upper panel of 
Fig.~\ref{fig:CRs_and_Biases_summary}. Although limited to only one point in the $(\eta,\chiPN)$ 
parameter space, this result suggests that the reduced-spin inspiral model is sufficiently accurate 
for parameter estimation from GW observations in aLIGO and AdV.

We then consider the hybrid PN-NR inspiral-merger-ringdown waveforms (with constant $\chiIMR = 0.45$), 
and compare them against the IMRPhenomC model. Our first observation is that, while the IMRPhenomC 
model performs well for detection purposes, with all fitting factors $>$0.99
 (see 
Fig.~\ref{fig:plots_IMRPhenomC_TaylorF2_hyb_FFs}), artifacts in the construction of the waveform model 
cause a significant systematic bias even for the equal-spin waveform that should be reproduced by the
model with $\chiIMR = 0.45$. This problem can be removed in future by producing a model calibrated against
NR waveforms across a larger volume of parameter space, and such work is already underway.
The bias we observe at low masses (see Figs.~\ref{fig:plots_IMRPhenomC_TaylorF2_hyb_mass_bias}, 
\ref{fig:plots_IMRPhenomC_TaylorF2_hyb_mass_ratio_bias}, and 
\ref{fig:plots_IMRPhenomC_TaylorF2_hyb_spin_bias}) is consistent with our choice of $\chiIMR$ as
our single-effective-spin parameter, and suggests that future phenomenological models should be 
parameterized instead with $\chiPN$. 

Our main observation from these results is that at intermediate masses (around 50\,$M_\odot$) 
the \emph{spread} in recovered spin values is far larger
than the statistical uncertainty in $\chiIMR$, even at an SNR of 10, which is close to the detection 
threshold. This can be seen in Figs.~\ref{fig:plots_IMRPhenomC_TaylorF2_hyb_spin_bias}
and \ref{fig:plots_IMRPhenomC_self_CRs_50M_SNR20}, and is summarized here in the lower panel of 
Fig.~\ref{fig:CRs_and_Biases_summary}, which shows the deviation in the parameter measurement from the 
value obtained for the reference $\chi_1 = \chi_2 = 0.45$ waveform.

The small parameter bias due to the single-effective-spin approximation at low masses implies that
the approximation holds well at these masses, and therefore that \emph{it will be difficult to measure the 
component spins} in low-mass binaries. This is the unfortunate corollary of the validity of the 
single-effective-spin approximation:
GW searches can be more efficient, but parameter estimation is less accurate. 

Since we observe large parameter biases at intermediate
masses, this implies that in these cases we \emph{may be} able to measure the individual spins.
It is quite likely that there is a strong degeneracy between the two spins and the mass ratio at 
all stages in the binary's evolution, but if we observe a waveform 
in which the early part (the inspiral) is characterized by $\chiPN$, and the late part (the ringdown) is 
characterized by the final spin, then it is likely that to describe the full waveform we require knowledge 
of both black-hole spins. If this is the case, then it follows that accurate measurements of both spins may 
be possible. This is an interesting topic for further work. 

This study analyzes only the $l=2, m=\pm 2$ modes of the gravitational wave signal.
The amplitude of higher order modes is sensitive to the mass-ratio, and thus could serve to break 
degeneracies between $\chi$ and $\eta$ in parameter estimation. However, it should be borne in mind 
that at the mass-ratios that this model is calibrated at the higher modes are all much weaker than the 
22-mode and at the low SNRs considered here we do not expect our results to change appreciably. 

Do these results imply that we should construct a two-spin non-precessing model? Not necessarily. 
A single-effective-spin model is sufficient for detection, and, following detection, if the source is a
black-hole-binary with total mass $\sim$50\,$M_\odot$, then we could perform addition simulations with 
varying individual black-hole spins, and produce a localized model for parameter estimation purposes. 
We note that although it is possible to produce sufficient waveforms to cover the non-precessing-binary
parameter space in time for the commissioning of aLIGO and AdV, a higher
priority may be to produce models that approximately cover the \emph{full} precessing-binary parameter
space, for which it would be more efficient to model the spins parallel to the orbital angular momentum
by only a single parameter. 

Further work is needed to verify the spread of the parameter biases in the spin in the IMR 
waveforms, using updated phenomenological waveform models, and studies across a larger
volume of the parameter space. In assessing our ability to measure either an effective single
spin parameter, or both black-hole spins, we also need to quantify the influence of harmonics
beyond the dominant $(\ell=2,|m|=2)$ modes.


\section{Acknowledgements} 

We thank S. Fairhurst, F. Ohme, H. Pfeiffer, S. Khan, B. Sathyaprakash, and P. Schmidt for useful discussions and 
comments.
M. Hannam was supported Science and Technology Facilities Council grants ST/H008438/1
and ST/I001085/1.
S. Husa was supported by
grant FPA-2007-60220 from the Spanish Ministry of Science and
the Spanish MICINN’s Consolider-Ingenio 2010 Programme under grant 
MultiDark CSD2009-00064.
M. P\"urrer thanks Caltech and the Universitat de les Illes Balears (UIB) for hospitality.
{\tt BAM} simulations were carried out at Advanced Research Computing 
(ARCCA) at Cardiff, and as part of the European PRACE petascale computing
initiative on the clusters Hermit, Curie and SuperMUC.


\bibliography{const_chi_IMR}

\begin{thebibliography}{72}
\expandafter\ifx\csname natexlab\endcsname\relax\def\natexlab#1{#1}\fi
\expandafter\ifx\csname bibnamefont\endcsname\relax
  \def\bibnamefont#1{#1}\fi
\expandafter\ifx\csname bibfnamefont\endcsname\relax
  \def\bibfnamefont#1{#1}\fi
\expandafter\ifx\csname citenamefont\endcsname\relax
  \def\citenamefont#1{#1}\fi
\expandafter\ifx\csname url\endcsname\relax
  \def\url#1{\texttt{#1}}\fi
\expandafter\ifx\csname urlprefix\endcsname\relax\def\urlprefix{URL }\fi
\providecommand{\bibinfo}[2]{#2}
\providecommand{\eprint}[2][]{\url{#2}}

\bibitem[{\citenamefont{Harry et~al.}(2010)}]{Harry:2010zz}
\bibinfo{author}{\bibfnamefont{G.~M.} \bibnamefont{Harry}} \bibnamefont{et~al.}
  (\bibinfo{collaboration}{LIGO Scientific Collaboration}),
  \bibinfo{journal}{Class.Quant.Grav.} \textbf{\bibinfo{volume}{27}},
  \bibinfo{pages}{084006} (\bibinfo{year}{2010}).

\bibitem[{\citenamefont{Acernese et~al.}(2009)}]{aVIRGO}
\bibinfo{author}{\bibfnamefont{F.}~\bibnamefont{Acernese}} \bibnamefont{et~al.}
  (\bibinfo{collaboration}{{The Virgo Collaboration}}),
  \emph{\bibinfo{title}{{Advanced Virgo Baseline Design}}}
  (\bibinfo{year}{2009}), \bibinfo{note}{{[Virgo Techincal Document
  VIR-0027A-09]}}.

\bibitem[{\citenamefont{Abadie et~al.}(2010{\natexlab{a}})}]{Abadie:2010cf}
\bibinfo{author}{\bibfnamefont{J.}~\bibnamefont{Abadie}} \bibnamefont{et~al.}
  (\bibinfo{collaboration}{LIGO Scientific}), \bibinfo{journal}{Class. Quant.
  Grav.} \textbf{\bibinfo{volume}{27}}, \bibinfo{pages}{173001}
  (\bibinfo{year}{2010}{\natexlab{a}}), \eprint{1003.2480}.

\bibitem[{\citenamefont{Sathyaprakash and Schutz}(2009)}]{Sathyaprakash:2009xs}
\bibinfo{author}{\bibfnamefont{B.}~\bibnamefont{Sathyaprakash}}
  \bibnamefont{and} \bibinfo{author}{\bibfnamefont{B.}~\bibnamefont{Schutz}},
  \bibinfo{journal}{Living Rev.Rel.} \textbf{\bibinfo{volume}{12}},
  \bibinfo{pages}{2} (\bibinfo{year}{2009}),
  \eprint{http://www.livingreviews.org/lrr-2009-2}.

\bibitem[{\citenamefont{Aasi et~al.}(2013{\natexlab{a}})}]{Aasi:2012rja}
\bibinfo{author}{\bibfnamefont{J.}~\bibnamefont{Aasi}} \bibnamefont{et~al.}
  (\bibinfo{collaboration}{LIGO Scientific Collaboration, Virgo
  Collaboration}), \bibinfo{journal}{Phys. Rev. D 87,}
  \textbf{\bibinfo{volume}{022002}}, \bibinfo{pages}{022002}
  (\bibinfo{year}{2013}{\natexlab{a}}), \eprint{1209.6533}.

\bibitem[{\citenamefont{Abadie et~al.}(2012)}]{Colaboration:2011np}
\bibinfo{author}{\bibfnamefont{J.}~\bibnamefont{Abadie}} \bibnamefont{et~al.}
  (\bibinfo{collaboration}{LIGO Collaboration, Virgo Collaboration}),
  \bibinfo{journal}{Phys.Rev.} \textbf{\bibinfo{volume}{D85}},
  \bibinfo{pages}{082002} (\bibinfo{year}{2012}), \eprint{1111.7314}.

\bibitem[{\citenamefont{Abadie et~al.}(2011)}]{Abadie:2011kd}
\bibinfo{author}{\bibfnamefont{J.}~\bibnamefont{Abadie}} \bibnamefont{et~al.}
  (\bibinfo{collaboration}{LIGO Scientific Collaboration, Virgo
  Collaboration}), \bibinfo{journal}{Phys.Rev.} \textbf{\bibinfo{volume}{D83}},
  \bibinfo{pages}{122005} (\bibinfo{year}{2011}), \eprint{1102.3781}.

\bibitem[{\citenamefont{Abadie et~al.}(2010{\natexlab{b}})}]{Abadie:2010yb}
\bibinfo{author}{\bibfnamefont{J.}~\bibnamefont{Abadie}} \bibnamefont{et~al.}
  (\bibinfo{collaboration}{LIGO Scientific Collaboration, Virgo
  Collaboration}), \bibinfo{journal}{Phys.Rev.} \textbf{\bibinfo{volume}{D82}},
  \bibinfo{pages}{102001} (\bibinfo{year}{2010}{\natexlab{b}}),
  \eprint{1005.4655}.

\bibitem[{\citenamefont{Ajith et~al.}(2011)\citenamefont{Ajith, Hannam, Husa,
  Chen, Bruegmann et~al.}}]{Ajith:2009bn}
\bibinfo{author}{\bibfnamefont{P.}~\bibnamefont{Ajith}},
  \bibinfo{author}{\bibfnamefont{M.}~\bibnamefont{Hannam}},
  \bibinfo{author}{\bibfnamefont{S.}~\bibnamefont{Husa}},
  \bibinfo{author}{\bibfnamefont{Y.}~\bibnamefont{Chen}},
  \bibinfo{author}{\bibfnamefont{B.}~\bibnamefont{Bruegmann}},
  \bibnamefont{et~al.}, \bibinfo{journal}{Phys.Rev.Lett.}
  \textbf{\bibinfo{volume}{106}}, \bibinfo{pages}{241101}
  (\bibinfo{year}{2011}), \eprint{0909.2867}.

\bibitem[{\citenamefont{Ajith}(2011)}]{Ajith:2011ec}
\bibinfo{author}{\bibfnamefont{P.}~\bibnamefont{Ajith}},
  \bibinfo{journal}{Phys.Rev.} \textbf{\bibinfo{volume}{D84}},
  \bibinfo{pages}{084037} (\bibinfo{year}{2011}), \eprint{1107.1267}.

\bibitem[{\citenamefont{Brown et~al.}(2012)\citenamefont{Brown, Harry,
  Lundgren, and Nitz}}]{Brown:2012qf}
\bibinfo{author}{\bibfnamefont{D.~A.} \bibnamefont{Brown}},
  \bibinfo{author}{\bibfnamefont{I.}~\bibnamefont{Harry}},
  \bibinfo{author}{\bibfnamefont{A.}~\bibnamefont{Lundgren}}, \bibnamefont{and}
  \bibinfo{author}{\bibfnamefont{A.~H.} \bibnamefont{Nitz}},
  \bibinfo{journal}{Phys.Rev.} \textbf{\bibinfo{volume}{D86}},
  \bibinfo{pages}{084017} (\bibinfo{year}{2012}), \eprint{1207.6406}.

\bibitem[{\citenamefont{Schmidt et~al.}(2012)\citenamefont{Schmidt, Hannam, and
  Husa}}]{Schmidt:2012rh}
\bibinfo{author}{\bibfnamefont{P.}~\bibnamefont{Schmidt}},
  \bibinfo{author}{\bibfnamefont{M.}~\bibnamefont{Hannam}}, \bibnamefont{and}
  \bibinfo{author}{\bibfnamefont{S.}~\bibnamefont{Husa}},
  \bibinfo{journal}{Phys.Rev.} \textbf{\bibinfo{volume}{D86}},
  \bibinfo{pages}{104063} (\bibinfo{year}{2012}), \eprint{1207.3088}.

\bibitem[{\citenamefont{Pekowsky et~al.}(2013)\citenamefont{Pekowsky,
  O'Shaughnessy, Healy, and Shoemaker}}]{Pekowsky:2013ska}
\bibinfo{author}{\bibfnamefont{L.}~\bibnamefont{Pekowsky}},
  \bibinfo{author}{\bibfnamefont{R.}~\bibnamefont{O'Shaughnessy}},
  \bibinfo{author}{\bibfnamefont{J.}~\bibnamefont{Healy}}, \bibnamefont{and}
  \bibinfo{author}{\bibfnamefont{D.}~\bibnamefont{Shoemaker}}
  (\bibinfo{year}{2013}), \eprint{1304.3176}.

\bibitem[{\citenamefont{Santamaria et~al.}(2010)}]{Santamaria:2010yb}
\bibinfo{author}{\bibfnamefont{L.}~\bibnamefont{Santamaria}}
  \bibnamefont{et~al.}, \bibinfo{journal}{Phys. Rev.}
  \textbf{\bibinfo{volume}{D82}}, \bibinfo{pages}{064016}
  (\bibinfo{year}{2010}), \eprint{1005.3306}.

\bibitem[{\citenamefont{Vaishnav et~al.}(2007)\citenamefont{Vaishnav, Hinder,
  Herrmann, and Shoemaker}}]{Vaishnav:2007nm}
\bibinfo{author}{\bibfnamefont{B.}~\bibnamefont{Vaishnav}},
  \bibinfo{author}{\bibfnamefont{I.}~\bibnamefont{Hinder}},
  \bibinfo{author}{\bibfnamefont{F.}~\bibnamefont{Herrmann}}, \bibnamefont{and}
  \bibinfo{author}{\bibfnamefont{D.}~\bibnamefont{Shoemaker}},
  \bibinfo{journal}{Phys. Rev.} \textbf{\bibinfo{volume}{D76}},
  \bibinfo{pages}{084020} (\bibinfo{year}{2007}), \eprint{0705.3829}.

\bibitem[{\citenamefont{Reisswig et~al.}(2009)}]{Reisswig:2009vc}
\bibinfo{author}{\bibfnamefont{C.}~\bibnamefont{Reisswig}}
  \bibnamefont{et~al.}, \bibinfo{journal}{Phys. Rev.}
  \textbf{\bibinfo{volume}{D80}}, \bibinfo{pages}{124026}
  (\bibinfo{year}{2009}), \eprint{0907.0462}.

\bibitem[{\citenamefont{Ajith et~al.}(2012{\natexlab{a}})\citenamefont{Ajith,
  Fotopoulos, Privitera, Neunzert, and Weinstein}}]{Ajith:2012mn}
\bibinfo{author}{\bibfnamefont{P.}~\bibnamefont{Ajith}},
  \bibinfo{author}{\bibfnamefont{N.}~\bibnamefont{Fotopoulos}},
  \bibinfo{author}{\bibfnamefont{S.}~\bibnamefont{Privitera}},
  \bibinfo{author}{\bibfnamefont{A.}~\bibnamefont{Neunzert}}, \bibnamefont{and}
  \bibinfo{author}{\bibfnamefont{A.}~\bibnamefont{Weinstein}}
  (\bibinfo{year}{2012}{\natexlab{a}}), \eprint{1210.6666}.

\bibitem[{\citenamefont{Ajith et~al.}(2007)}]{Ajith:2007qp}
\bibinfo{author}{\bibfnamefont{P.}~\bibnamefont{Ajith}} \bibnamefont{et~al.},
  \bibinfo{journal}{Class. Quant. Grav.} \textbf{\bibinfo{volume}{24}},
  \bibinfo{pages}{S689} (\bibinfo{year}{2007}), \eprint{0704.3764}.

\bibitem[{\citenamefont{Ajith et~al.}(2008)}]{Ajith:2007kx}
\bibinfo{author}{\bibfnamefont{P.}~\bibnamefont{Ajith}} \bibnamefont{et~al.},
  \bibinfo{journal}{Phys. Rev.} \textbf{\bibinfo{volume}{D77}},
  \bibinfo{pages}{104017} (\bibinfo{year}{2008}), \eprint{0710.2335}.

\bibitem[{\citenamefont{Ajith}(2008)}]{Ajith:2007xh}
\bibinfo{author}{\bibfnamefont{P.}~\bibnamefont{Ajith}},
  \bibinfo{journal}{Class. Quant. Grav.} \textbf{\bibinfo{volume}{25}},
  \bibinfo{pages}{114033} (\bibinfo{year}{2008}), \eprint{0712.0343}.

\bibitem[{\citenamefont{Hannam and Hawke}(2011)}]{Hannam:2009vt}
\bibinfo{author}{\bibfnamefont{M.}~\bibnamefont{Hannam}} \bibnamefont{and}
  \bibinfo{author}{\bibfnamefont{I.}~\bibnamefont{Hawke}},
  \bibinfo{journal}{Gen.Rel.Grav.} \textbf{\bibinfo{volume}{43}},
  \bibinfo{pages}{465} (\bibinfo{year}{2011}), \eprint{0908.3139}.

\bibitem[{\citenamefont{Mroue et~al.}(2013)\citenamefont{Mroue, Scheel,
  Szilagyi, Pfeiffer, Boyle et~al.}}]{Mroue:2013xna}
\bibinfo{author}{\bibfnamefont{A.~H.} \bibnamefont{Mroue}},
  \bibinfo{author}{\bibfnamefont{M.~A.} \bibnamefont{Scheel}},
  \bibinfo{author}{\bibfnamefont{B.}~\bibnamefont{Szilagyi}},
  \bibinfo{author}{\bibfnamefont{H.~P.} \bibnamefont{Pfeiffer}},
  \bibinfo{author}{\bibfnamefont{M.}~\bibnamefont{Boyle}}, \bibnamefont{et~al.}
  (\bibinfo{year}{2013}), \eprint{1304.6077}.

\bibitem[{\citenamefont{Buonanno et~al.}(2008)\citenamefont{Buonanno, Kidder,
  and Lehner}}]{Buonanno:2007sv}
\bibinfo{author}{\bibfnamefont{A.}~\bibnamefont{Buonanno}},
  \bibinfo{author}{\bibfnamefont{L.~E.} \bibnamefont{Kidder}},
  \bibnamefont{and} \bibinfo{author}{\bibfnamefont{L.}~\bibnamefont{Lehner}},
  \bibinfo{journal}{Phys. Rev.} \textbf{\bibinfo{volume}{D77}},
  \bibinfo{pages}{026004} (\bibinfo{year}{2008}).

\bibitem[{\citenamefont{Rezzolla et~al.}(2008{\natexlab{a}})}]{Rezzolla:2007rd}
\bibinfo{author}{\bibfnamefont{L.}~\bibnamefont{Rezzolla}}
  \bibnamefont{et~al.}, \bibinfo{journal}{Astrophys. J.}
  \textbf{\bibinfo{volume}{674}}, \bibinfo{pages}{L29}
  (\bibinfo{year}{2008}{\natexlab{a}}), \eprint{0710.3345}.

\bibitem[{\citenamefont{Tichy and Marronetti}(2008)}]{Tichy:2008du}
\bibinfo{author}{\bibfnamefont{W.}~\bibnamefont{Tichy}} \bibnamefont{and}
  \bibinfo{author}{\bibfnamefont{P.}~\bibnamefont{Marronetti}},
  \bibinfo{journal}{Phys.Rev.} \textbf{\bibinfo{volume}{D78}},
  \bibinfo{pages}{081501} (\bibinfo{year}{2008}), \eprint{0807.2985}.

\bibitem[{\citenamefont{Lousto et~al.}(2010)\citenamefont{Lousto, Campanelli,
  and Zlochower}}]{Lousto:2009mf}
\bibinfo{author}{\bibfnamefont{C.~O.} \bibnamefont{Lousto}},
  \bibinfo{author}{\bibfnamefont{M.}~\bibnamefont{Campanelli}},
  \bibnamefont{and}
  \bibinfo{author}{\bibfnamefont{Y.}~\bibnamefont{Zlochower}},
  \bibinfo{journal}{Class. Quant. Grav.} \textbf{\bibinfo{volume}{27}},
  \bibinfo{pages}{114006} (\bibinfo{year}{2010}), \eprint{0904.3541}.

\bibitem[{\citenamefont{Hannam et~al.}(2010)\citenamefont{Hannam, Husa, Ohme,
  and Ajith}}]{Hannam:2010ky}
\bibinfo{author}{\bibfnamefont{M.}~\bibnamefont{Hannam}},
  \bibinfo{author}{\bibfnamefont{S.}~\bibnamefont{Husa}},
  \bibinfo{author}{\bibfnamefont{F.}~\bibnamefont{Ohme}}, \bibnamefont{and}
  \bibinfo{author}{\bibfnamefont{P.}~\bibnamefont{Ajith}},
  \bibinfo{journal}{Phys.Rev.} \textbf{\bibinfo{volume}{D82}},
  \bibinfo{pages}{124052} (\bibinfo{year}{2010}), \eprint{1008.2961}.

\bibitem[{\citenamefont{MacDonald et~al.}(2011)\citenamefont{MacDonald,
  Nissanke, Pfeiffer, and Pfeiffer}}]{MacDonald:2011ne}
\bibinfo{author}{\bibfnamefont{I.}~\bibnamefont{MacDonald}},
  \bibinfo{author}{\bibfnamefont{S.}~\bibnamefont{Nissanke}},
  \bibinfo{author}{\bibfnamefont{H.~P.} \bibnamefont{Pfeiffer}},
  \bibnamefont{and} \bibinfo{author}{\bibfnamefont{H.~P.}
  \bibnamefont{Pfeiffer}}, \bibinfo{journal}{Class.Quant.Grav.}
  \textbf{\bibinfo{volume}{28}}, \bibinfo{pages}{134002}
  (\bibinfo{year}{2011}), \eprint{1102.5128}.

\bibitem[{\citenamefont{Boyle}(2011)}]{Boyle:2011dy}
\bibinfo{author}{\bibfnamefont{M.}~\bibnamefont{Boyle}},
  \bibinfo{journal}{Phys.Rev.} \textbf{\bibinfo{volume}{D84}},
  \bibinfo{pages}{064013} (\bibinfo{year}{2011}), \eprint{1103.5088}.

\bibitem[{\citenamefont{Ohme et~al.}(2011)\citenamefont{Ohme, Hannam, and
  Husa}}]{Ohme:2011zm}
\bibinfo{author}{\bibfnamefont{F.}~\bibnamefont{Ohme}},
  \bibinfo{author}{\bibfnamefont{M.}~\bibnamefont{Hannam}}, \bibnamefont{and}
  \bibinfo{author}{\bibfnamefont{S.}~\bibnamefont{Husa}},
  \bibinfo{journal}{Phys.Rev.} \textbf{\bibinfo{volume}{D84}},
  \bibinfo{pages}{064029} (\bibinfo{year}{2011}), \eprint{1107.0996}.

\bibitem[{\citenamefont{Poisson and Will}(1995)}]{Poisson:1995ef}
\bibinfo{author}{\bibfnamefont{E.}~\bibnamefont{Poisson}} \bibnamefont{and}
  \bibinfo{author}{\bibfnamefont{C.~M.} \bibnamefont{Will}},
  \bibinfo{journal}{Phys.Rev.} \textbf{\bibinfo{volume}{D52}},
  \bibinfo{pages}{848} (\bibinfo{year}{1995}), \eprint{gr-qc/9502040}.

\bibitem[{LAL()}]{LAL}
\emph{\bibinfo{title}{{LSC} algorithm library {(LAL)}}},
  \urlprefix\url{http://www.lsc-group.phys.uwm.edu/lal}.

\bibitem[{\citenamefont{Sathyaprakash and
  Dhurandhar}(1991)}]{Sathyaprakash:1991mt}
\bibinfo{author}{\bibfnamefont{B.}~\bibnamefont{Sathyaprakash}}
  \bibnamefont{and}
  \bibinfo{author}{\bibfnamefont{S.}~\bibnamefont{Dhurandhar}},
  \bibinfo{journal}{Phys.Rev.} \textbf{\bibinfo{volume}{D44}},
  \bibinfo{pages}{3819} (\bibinfo{year}{1991}).

\bibitem[{\citenamefont{Cutler and
  Flanagan}(1994{\natexlab{a}})}]{Cutler:1994ys}
\bibinfo{author}{\bibfnamefont{C.}~\bibnamefont{Cutler}} \bibnamefont{and}
  \bibinfo{author}{\bibfnamefont{E.~E.} \bibnamefont{Flanagan}},
  \bibinfo{journal}{Phys.Rev.} \textbf{\bibinfo{volume}{D49}},
  \bibinfo{pages}{2658} (\bibinfo{year}{1994}{\natexlab{a}}).

\bibitem[{\citenamefont{Droz et~al.}(1999)\citenamefont{Droz, Knapp, Poisson,
  and Owen}}]{Droz:1999qx}
\bibinfo{author}{\bibfnamefont{S.}~\bibnamefont{Droz}},
  \bibinfo{author}{\bibfnamefont{D.~J.} \bibnamefont{Knapp}},
  \bibinfo{author}{\bibfnamefont{E.}~\bibnamefont{Poisson}}, \bibnamefont{and}
  \bibinfo{author}{\bibfnamefont{B.~J.} \bibnamefont{Owen}},
  \bibinfo{journal}{Phys. Rev. D} \textbf{\bibinfo{volume}{59}},
  \bibinfo{pages}{124016} (\bibinfo{year}{1999}), \eprint{gr-qc/9901076}.

\bibitem[{\citenamefont{Ajith et~al.}(2012{\natexlab{b}})\citenamefont{Ajith,
  Boyle, Brown, Brugmann, Buchman et~al.}}]{Ajith:2012az}
\bibinfo{author}{\bibfnamefont{P.}~\bibnamefont{Ajith}},
  \bibinfo{author}{\bibfnamefont{M.}~\bibnamefont{Boyle}},
  \bibinfo{author}{\bibfnamefont{D.~A.} \bibnamefont{Brown}},
  \bibinfo{author}{\bibfnamefont{B.}~\bibnamefont{Brugmann}},
  \bibinfo{author}{\bibfnamefont{L.~T.} \bibnamefont{Buchman}},
  \bibnamefont{et~al.}, \bibinfo{journal}{Class.Quant.Grav.}
  \textbf{\bibinfo{volume}{29}}, \bibinfo{pages}{124001}
  (\bibinfo{year}{2012}{\natexlab{b}}), \eprint{1201.5319}.

\bibitem[{\citenamefont{Cutler and Flanagan}(1994{\natexlab{b}})}]{Cutler94}
\bibinfo{author}{\bibfnamefont{C.}~\bibnamefont{Cutler}} \bibnamefont{and}
  \bibinfo{author}{\bibfnamefont{E.~E.} \bibnamefont{Flanagan}},
  \bibinfo{journal}{Phys. Rev. D} \textbf{\bibinfo{volume}{49}},
  \bibinfo{pages}{2658} (\bibinfo{year}{1994}{\natexlab{b}}).

\bibitem[{\citenamefont{Abbott et~al.}(2009)}]{Abbott:2007kv}
\bibinfo{author}{\bibfnamefont{B.}~\bibnamefont{Abbott}} \bibnamefont{et~al.}
  (\bibinfo{collaboration}{LIGO Scientific}), \bibinfo{journal}{Rept. Prog.
  Phys.} \textbf{\bibinfo{volume}{72}}, \bibinfo{pages}{076901}
  (\bibinfo{year}{2009}), \eprint{0711.3041}.

\bibitem[{\citenamefont{Shoemaker}(2009)}]{Shoemaker:aLIGO}
\bibinfo{author}{\bibfnamefont{D.}~\bibnamefont{Shoemaker}}
  (\bibinfo{collaboration}{the Advanced LIGO Team}),
  \emph{\bibinfo{title}{{Advanced LIGO Reference Design}}}
  (\bibinfo{year}{2009}), \bibinfo{note}{{[LIGO-M060056]}}.

\bibitem[{\citenamefont{{Harry} and {the LIGO Scientific
  Collaboration}}(2010)}]{2010CQGra..27h4006H}
\bibinfo{author}{\bibfnamefont{G.~M.} \bibnamefont{{Harry}}} \bibnamefont{and}
  \bibinfo{author}{\bibnamefont{{the LIGO Scientific Collaboration}}},
  \bibinfo{journal}{Class. Quant. Grav.} \textbf{\bibinfo{volume}{27}},
  \bibinfo{pages}{084006} (\bibinfo{year}{2010}).

\bibitem[{\citenamefont{Aasi et~al.}(2013{\natexlab{b}})}]{Aasi:2013wya}
\bibinfo{author}{\bibfnamefont{J.}~\bibnamefont{Aasi}} \bibnamefont{et~al.}
  (\bibinfo{collaboration}{LIGO Scientific Collaboration, Virgo Collaboration})
  (\bibinfo{year}{2013}{\natexlab{b}}), \eprint{1304.0670}.

\bibitem[{\citenamefont{{The LIGO Scientific Collaboration}}(2009)}]{T0900288}
\bibinfo{author}{\bibnamefont{{The LIGO Scientific Collaboration}}},
  \bibinfo{type}{Tech. Rep.} \bibinfo{number}{{LIGO}-T0900288-v3},
  \bibinfo{institution}{{LIGO} Project} (\bibinfo{year}{2009}),
  \urlprefix\url{https://dcc.ligo.org/DocDB/0002/T0900288/003/AdvLIGO%20noise%20curves.pdf}.

\bibitem[{\citenamefont{Babak et~al.}(2013)\citenamefont{Babak, Biswas, Brady,
  Brown, Cannon et~al.}}]{Babak:2012zx}
\bibinfo{author}{\bibfnamefont{S.}~\bibnamefont{Babak}},
  \bibinfo{author}{\bibfnamefont{R.}~\bibnamefont{Biswas}},
  \bibinfo{author}{\bibfnamefont{P.}~\bibnamefont{Brady}},
  \bibinfo{author}{\bibfnamefont{D.}~\bibnamefont{Brown}},
  \bibinfo{author}{\bibfnamefont{K.}~\bibnamefont{Cannon}},
  \bibnamefont{et~al.}, \bibinfo{journal}{Phys.Rev.}
  \textbf{\bibinfo{volume}{D87}}, \bibinfo{pages}{024033}
  (\bibinfo{year}{2013}), \eprint{1208.3491}.

\bibitem[{\citenamefont{Finn and Chernoff}(1993)}]{Finn:1992xs}
\bibinfo{author}{\bibfnamefont{L.~S.} \bibnamefont{Finn}} \bibnamefont{and}
  \bibinfo{author}{\bibfnamefont{D.~F.} \bibnamefont{Chernoff}},
  \bibinfo{journal}{Phys.Rev.} \textbf{\bibinfo{volume}{D47}},
  \bibinfo{pages}{2198} (\bibinfo{year}{1993}), \eprint{gr-qc/9301003}.

\bibitem[{\citenamefont{Arun et~al.}(2005)\citenamefont{Arun, Iyer,
  Sathyaprakash, and Sundararajan}}]{Arun:2004hn}
\bibinfo{author}{\bibfnamefont{K.}~\bibnamefont{Arun}},
  \bibinfo{author}{\bibfnamefont{B.~R.} \bibnamefont{Iyer}},
  \bibinfo{author}{\bibfnamefont{B.}~\bibnamefont{Sathyaprakash}},
  \bibnamefont{and} \bibinfo{author}{\bibfnamefont{P.~A.}
  \bibnamefont{Sundararajan}}, \bibinfo{journal}{Phys.Rev.}
  \textbf{\bibinfo{volume}{D71}}, \bibinfo{pages}{084008}
  (\bibinfo{year}{2005}).

\bibitem[{\citenamefont{{van der Sluys}
  et~al.}(2008{\natexlab{a}})}]{Sluys:2008a}
\bibinfo{author}{\bibfnamefont{M.}~\bibnamefont{{van der Sluys}}}
  \bibnamefont{et~al.}, \bibinfo{journal}{Classical and Quantum Gravity}
  \textbf{\bibinfo{volume}{25}}, \bibinfo{pages}{184011}
  (\bibinfo{year}{2008}{\natexlab{a}}), \eprint{0805.1689}.

\bibitem[{\citenamefont{{van der Sluys}
  et~al.}(2008{\natexlab{b}})}]{Sluys:2008b}
\bibinfo{author}{\bibfnamefont{M.~V.} \bibnamefont{{van der Sluys}}}
  \bibnamefont{et~al.}, \bibinfo{journal}{The Astrophysical Journal Letters}
  \textbf{\bibinfo{volume}{688}}, \bibinfo{pages}{L61}
  (\bibinfo{year}{2008}{\natexlab{b}}), \eprint{0710.1897}.

\bibitem[{\citenamefont{{Veitch} and {Vecchio}}(2010)}]{Veitch:2010}
\bibinfo{author}{\bibfnamefont{J.}~\bibnamefont{{Veitch}}} \bibnamefont{and}
  \bibinfo{author}{\bibfnamefont{A.}~\bibnamefont{{Vecchio}}},
  \bibinfo{journal}{\prd} \textbf{\bibinfo{volume}{81}},
  \bibinfo{pages}{062003} (\bibinfo{year}{2010}), \eprint{0911.3820}.

\bibitem[{\citenamefont{{Feroz} et~al.}(2009)\citenamefont{{Feroz}, {Hobson},
  and {Bridges}}}]{Feroz:2009}
\bibinfo{author}{\bibfnamefont{F.}~\bibnamefont{{Feroz}}},
  \bibinfo{author}{\bibfnamefont{M.~P.} \bibnamefont{{Hobson}}},
  \bibnamefont{and}
  \bibinfo{author}{\bibfnamefont{M.}~\bibnamefont{{Bridges}}},
  \bibinfo{journal}{Monthly Notices of the Royal Astronomical Society}
  \textbf{\bibinfo{volume}{398}}, \bibinfo{pages}{1601} (\bibinfo{year}{2009}),
  \eprint{0809.3437}.

\bibitem[{\citenamefont{Aasi et~al.}(2013{\natexlab{c}})}]{Aasi:2013jjl}
\bibinfo{author}{\bibfnamefont{J.}~\bibnamefont{Aasi}} \bibnamefont{et~al.}
  (\bibinfo{collaboration}{LIGO Collaboration, Virgo Collaboration})
  (\bibinfo{year}{2013}{\natexlab{c}}), \eprint{1304.1775}.

\bibitem[{\citenamefont{Baird et~al.}(2013)\citenamefont{Baird, Fairhurst,
  Hannam, and Murphy}}]{Baird:2012cu}
\bibinfo{author}{\bibfnamefont{E.}~\bibnamefont{Baird}},
  \bibinfo{author}{\bibfnamefont{S.}~\bibnamefont{Fairhurst}},
  \bibinfo{author}{\bibfnamefont{M.}~\bibnamefont{Hannam}}, \bibnamefont{and}
  \bibinfo{author}{\bibfnamefont{P.}~\bibnamefont{Murphy}},
  \bibinfo{journal}{Phys.Rev.} \textbf{\bibinfo{volume}{D87}},
  \bibinfo{pages}{024035} (\bibinfo{year}{2013}), \eprint{1211.0546}.

\bibitem[{\citenamefont{Nelder and Mead}(1965)}]{Nelder:1965zz}
\bibinfo{author}{\bibfnamefont{J.}~\bibnamefont{Nelder}} \bibnamefont{and}
  \bibinfo{author}{\bibfnamefont{R.}~\bibnamefont{Mead}},
  \bibinfo{journal}{Comput.J.} \textbf{\bibinfo{volume}{7}},
  \bibinfo{pages}{308} (\bibinfo{year}{1965}).

\bibitem[{\citenamefont{Ohme et~al.}(2013)\citenamefont{Ohme, Nielsen, Keppel,
  and Lundgren}}]{Ohme:2013nsa}
\bibinfo{author}{\bibfnamefont{F.}~\bibnamefont{Ohme}},
  \bibinfo{author}{\bibfnamefont{A.~B.} \bibnamefont{Nielsen}},
  \bibinfo{author}{\bibfnamefont{D.}~\bibnamefont{Keppel}}, \bibnamefont{and}
  \bibinfo{author}{\bibfnamefont{A.}~\bibnamefont{Lundgren}}
  (\bibinfo{year}{2013}), \eprint{1304.7017}.

\bibitem[{\citenamefont{Tanaka and Tagoshi}(2000)}]{Tanaka:2000xy}
\bibinfo{author}{\bibfnamefont{T.}~\bibnamefont{Tanaka}} \bibnamefont{and}
  \bibinfo{author}{\bibfnamefont{H.}~\bibnamefont{Tagoshi}},
  \bibinfo{journal}{Phys.Rev.} \textbf{\bibinfo{volume}{D62}},
  \bibinfo{pages}{082001} (\bibinfo{year}{2000}), \eprint{gr-qc/0001090}.

\bibitem[{\citenamefont{Pai and Arun}(2013)}]{Pai:2012mv}
\bibinfo{author}{\bibfnamefont{A.}~\bibnamefont{Pai}} \bibnamefont{and}
  \bibinfo{author}{\bibfnamefont{K.}~\bibnamefont{Arun}},
  \bibinfo{journal}{Class.Quant.Grav.} \textbf{\bibinfo{volume}{30}},
  \bibinfo{pages}{025011} (\bibinfo{year}{2013}), \eprint{1207.1943}.

\bibitem[{\citenamefont{P{\"u}rrer et~al.}(2012)\citenamefont{P{\"u}rrer, Husa,
  and Hannam}}]{Purrer:2012wy}
\bibinfo{author}{\bibfnamefont{M.}~\bibnamefont{P{\"u}rrer}},
  \bibinfo{author}{\bibfnamefont{S.}~\bibnamefont{Husa}}, \bibnamefont{and}
  \bibinfo{author}{\bibfnamefont{M.}~\bibnamefont{Hannam}},
  \bibinfo{journal}{Phys.Rev.} \textbf{\bibinfo{volume}{D85}},
  \bibinfo{pages}{124051} (\bibinfo{year}{2012}), \eprint{1203.4258}.

\bibitem[{\citenamefont{Br{\"u}gmann et~al.}(2008)}]{Brugmann:2008zz}
\bibinfo{author}{\bibfnamefont{B.}~\bibnamefont{Br{\"u}gmann}}
  \bibnamefont{et~al.}, \bibinfo{journal}{Phys. Rev.}
  \textbf{\bibinfo{volume}{D77}}, \bibinfo{pages}{024027}
  (\bibinfo{year}{2008}), \eprint{gr-qc/0610128}.

\bibitem[{\citenamefont{Husa et~al.}(2008)\citenamefont{Husa, Gonz{\'a}lez,
  Hannam, Br{\"u}gmann, and Sperhake}}]{Husa:2007hp}
\bibinfo{author}{\bibfnamefont{S.}~\bibnamefont{Husa}},
  \bibinfo{author}{\bibfnamefont{J.~A.} \bibnamefont{Gonz{\'a}lez}},
  \bibinfo{author}{\bibfnamefont{M.}~\bibnamefont{Hannam}},
  \bibinfo{author}{\bibfnamefont{B.}~\bibnamefont{Br{\"u}gmann}},
  \bibnamefont{and} \bibinfo{author}{\bibfnamefont{U.}~\bibnamefont{Sperhake}},
  \bibinfo{journal}{Class. Quant. Grav.} \textbf{\bibinfo{volume}{25}},
  \bibinfo{pages}{105006} (\bibinfo{year}{2008}), \eprint{0706.0740}.

\bibitem[{\citenamefont{Brandt and Br{\"u}gmann}(1997)}]{Brandt:1997tf}
\bibinfo{author}{\bibfnamefont{S.}~\bibnamefont{Brandt}} \bibnamefont{and}
  \bibinfo{author}{\bibfnamefont{B.}~\bibnamefont{Br{\"u}gmann}},
  \bibinfo{journal}{Phys. Rev. Lett.} \textbf{\bibinfo{volume}{78}},
  \bibinfo{pages}{3606} (\bibinfo{year}{1997}), \eprint{gr-qc/9703066}.

\bibitem[{\citenamefont{Bowen and York}(1980)}]{Bowen:1980yu}
\bibinfo{author}{\bibfnamefont{J.~M.} \bibnamefont{Bowen}} \bibnamefont{and}
  \bibinfo{author}{\bibfnamefont{J.~W.} \bibnamefont{York},
  \bibfnamefont{Jr.}}, \bibinfo{journal}{Phys. Rev.}
  \textbf{\bibinfo{volume}{D21}}, \bibinfo{pages}{2047} (\bibinfo{year}{1980}).

\bibitem[{\citenamefont{Ansorg et~al.}(2004)\citenamefont{Ansorg, Br{\"u}gmann,
  and Tichy}}]{Ansorg:2004ds}
\bibinfo{author}{\bibfnamefont{M.}~\bibnamefont{Ansorg}},
  \bibinfo{author}{\bibfnamefont{B.}~\bibnamefont{Br{\"u}gmann}},
  \bibnamefont{and} \bibinfo{author}{\bibfnamefont{W.}~\bibnamefont{Tichy}},
  \bibinfo{journal}{Phys. Rev.} \textbf{\bibinfo{volume}{D70}},
  \bibinfo{pages}{064011} (\bibinfo{year}{2004}), \eprint{gr-qc/0404056}.

\bibitem[{\citenamefont{Campanelli et~al.}(2006)\citenamefont{Campanelli,
  Lousto, Marronetti, and Zlochower}}]{Campanelli:2005dd}
\bibinfo{author}{\bibfnamefont{M.}~\bibnamefont{Campanelli}},
  \bibinfo{author}{\bibfnamefont{C.~O.} \bibnamefont{Lousto}},
  \bibinfo{author}{\bibfnamefont{P.}~\bibnamefont{Marronetti}},
  \bibnamefont{and}
  \bibinfo{author}{\bibfnamefont{Y.}~\bibnamefont{Zlochower}},
  \bibinfo{journal}{Phys. Rev. Lett.} \textbf{\bibinfo{volume}{96}},
  \bibinfo{pages}{111101} (\bibinfo{year}{2006}), \eprint{gr-qc/0511048}.

\bibitem[{\citenamefont{Baker et~al.}(2006)\citenamefont{Baker, Centrella,
  Choi, Koppitz, and van Meter}}]{Baker:2005vv}
\bibinfo{author}{\bibfnamefont{J.~G.} \bibnamefont{Baker}},
  \bibinfo{author}{\bibfnamefont{J.}~\bibnamefont{Centrella}},
  \bibinfo{author}{\bibfnamefont{D.-I.} \bibnamefont{Choi}},
  \bibinfo{author}{\bibfnamefont{M.}~\bibnamefont{Koppitz}}, \bibnamefont{and}
  \bibinfo{author}{\bibfnamefont{J.}~\bibnamefont{van Meter}},
  \bibinfo{journal}{Phys. Rev. Lett.} \textbf{\bibinfo{volume}{96}},
  \bibinfo{pages}{111102} (\bibinfo{year}{2006}), \eprint{gr-qc/0511103}.

\bibitem[{\citenamefont{Hannam et~al.}(2007)\citenamefont{Hannam, Husa,
  Pollney, Br{\"u}gmann, and {\'O Murchadha}}}]{Hannam:2006vv}
\bibinfo{author}{\bibfnamefont{M.}~\bibnamefont{Hannam}},
  \bibinfo{author}{\bibfnamefont{S.}~\bibnamefont{Husa}},
  \bibinfo{author}{\bibfnamefont{D.}~\bibnamefont{Pollney}},
  \bibinfo{author}{\bibfnamefont{B.}~\bibnamefont{Br{\"u}gmann}},
  \bibnamefont{and} \bibinfo{author}{\bibfnamefont{N.}~\bibnamefont{{\'O
  Murchadha}}}, \bibinfo{journal}{Phys. Rev. Lett.}
  \textbf{\bibinfo{volume}{99}}, \bibinfo{pages}{241102}
  (\bibinfo{year}{2007}), \eprint{gr-qc/0606099}.

\bibitem[{\citenamefont{Shibata and Nakamura}(1995)}]{Shibata:1995we}
\bibinfo{author}{\bibfnamefont{M.}~\bibnamefont{Shibata}} \bibnamefont{and}
  \bibinfo{author}{\bibfnamefont{T.}~\bibnamefont{Nakamura}},
  \bibinfo{journal}{Phys. Rev.} \textbf{\bibinfo{volume}{D52}},
  \bibinfo{pages}{5428} (\bibinfo{year}{1995}).

\bibitem[{\citenamefont{Baumgarte and Shapiro}(1999)}]{Baumgarte:1998te}
\bibinfo{author}{\bibfnamefont{T.~W.} \bibnamefont{Baumgarte}}
  \bibnamefont{and} \bibinfo{author}{\bibfnamefont{S.~L.}
  \bibnamefont{Shapiro}}, \bibinfo{journal}{Phys. Rev.}
  \textbf{\bibinfo{volume}{D59}}, \bibinfo{pages}{024007}
  (\bibinfo{year}{1999}), \eprint{gr-qc/9810065}.

\bibitem[{\citenamefont{Husa et~al.}(2013)}]{Husa:2013}
\bibinfo{author}{\bibfnamefont{S.}~\bibnamefont{Husa}} \bibnamefont{et~al.}
  (\bibinfo{year}{2013}), \bibinfo{note}{in preparation}.

\bibitem[{\citenamefont{McKechan et~al.}(2010)\citenamefont{McKechan, Robinson,
  and Sathyaprakash}}]{McKechan:2010kp}
\bibinfo{author}{\bibfnamefont{D.}~\bibnamefont{McKechan}},
  \bibinfo{author}{\bibfnamefont{C.}~\bibnamefont{Robinson}}, \bibnamefont{and}
  \bibinfo{author}{\bibfnamefont{B.}~\bibnamefont{Sathyaprakash}},
  \bibinfo{journal}{Class.Quant.Grav.} \textbf{\bibinfo{volume}{27}},
  \bibinfo{pages}{084020} (\bibinfo{year}{2010}), \eprint{1003.2939}.

\bibitem[{\citenamefont{Lindblom et~al.}(2008)\citenamefont{Lindblom, Owen, and
  Brown}}]{Lindblom:2008cm}
\bibinfo{author}{\bibfnamefont{L.}~\bibnamefont{Lindblom}},
  \bibinfo{author}{\bibfnamefont{B.~J.} \bibnamefont{Owen}}, \bibnamefont{and}
  \bibinfo{author}{\bibfnamefont{D.~A.} \bibnamefont{Brown}},
  \bibinfo{journal}{Phys. Rev.} \textbf{\bibinfo{volume}{D78}},
  \bibinfo{pages}{124020} (\bibinfo{year}{2008}), \eprint{0809.3844}.

\bibitem[{\citenamefont{Hannam et~al.}(2013)\citenamefont{Hannam, Brown,
  Fairhurst, Fryer, and Harry}}]{Hannam:2013uu}
\bibinfo{author}{\bibfnamefont{M.}~\bibnamefont{Hannam}},
  \bibinfo{author}{\bibfnamefont{D.~A.} \bibnamefont{Brown}},
  \bibinfo{author}{\bibfnamefont{S.}~\bibnamefont{Fairhurst}},
  \bibinfo{author}{\bibfnamefont{C.~L.} \bibnamefont{Fryer}}, \bibnamefont{and}
  \bibinfo{author}{\bibfnamefont{I.~W.} \bibnamefont{Harry}},
  \bibinfo{journal}{Astrophys.J.} \textbf{\bibinfo{volume}{766}},
  \bibinfo{pages}{L14} (\bibinfo{year}{2013}), \eprint{1301.5616}.

\bibitem[{\citenamefont{Barausse and Rezzolla}(2009)}]{Barausse:2009uz}
\bibinfo{author}{\bibfnamefont{E.}~\bibnamefont{Barausse}} \bibnamefont{and}
  \bibinfo{author}{\bibfnamefont{L.}~\bibnamefont{Rezzolla}},
  \bibinfo{journal}{Astrophys.J.} \textbf{\bibinfo{volume}{704}},
  \bibinfo{pages}{L40} (\bibinfo{year}{2009}), \eprint{0904.2577}.

\bibitem[{\citenamefont{Rezzolla
  et~al.}(2008{\natexlab{b}})\citenamefont{Rezzolla, Barausse, Dorband,
  Pollney, Reisswig, Seiler, and Husa}}]{Rezzolla:2007rz}
\bibinfo{author}{\bibfnamefont{L.}~\bibnamefont{Rezzolla}},
  \bibinfo{author}{\bibfnamefont{E.}~\bibnamefont{Barausse}},
  \bibinfo{author}{\bibfnamefont{E.~N.} \bibnamefont{Dorband}},
  \bibinfo{author}{\bibfnamefont{D.}~\bibnamefont{Pollney}},
  \bibinfo{author}{\bibfnamefont{C.}~\bibnamefont{Reisswig}},
  \bibinfo{author}{\bibfnamefont{J.}~\bibnamefont{Seiler}}, \bibnamefont{and}
  \bibinfo{author}{\bibfnamefont{S.}~\bibnamefont{Husa}},
  \bibinfo{journal}{Phys. Rev.} \textbf{\bibinfo{volume}{D78}},
  \bibinfo{pages}{044002} (\bibinfo{year}{2008}{\natexlab{b}}),
  \eprint{0712.3541}.

\end{thebibliography}

\end{document}